\documentclass[sigplan,screen,nonacm]{acmart}
\renewcommand\footnotetextcopyrightpermission[1]{}
\usepackage{tikz}
\usepackage{amsmath}
\usepackage{graphicx}
\usepackage{booktabs}
\usepackage{hyperref}
\usepackage{enumitem}
\usepackage{algorithm}
\usepackage{algpseudocode}
\usepackage{xspace}
\usepackage{soul}
\usepackage{verbatim}
\usepackage{multirow}
\usepackage{pifont}

\newcommand\blfootnote[1]{%
  \begingroup
  \renewcommand\thefootnote{}\footnote{#1}%
  \addtocounter{footnote}{-1}%
  \endgroup
}

\newcommand{\system}{{\textsc{ForeMoE}}\xspace}
\newcommand{\mysubsubsection}[1]{{\vspace{0.25em}\noindent{\textbf{#1\xspace}}}}

\newcommand{\lhy}[1]{\textcolor{red}{(lhy: #1)}}

\newcommand{\myvspace}[1]{}

\usepackage{filecontents}

\begin{document}

\pagestyle{plain}

\date{}

\title{\Large \bf Harnessing Routing Foresight for Micro-step-level MoE load balancing in RL Post-training}

\author{\rm{Yuming Zhou}$^{*\dag}$ \hspace{1.2em} \rm{Haoyang Li}$^{*\dag}$ \hspace{1.2em} \rm{Sheng Lin}$^{\dag}$ \hspace{1.2em} \rm{Yanfeng Zhao}$^{\ddag}$  \hspace{1.2em} \\
 \rm{Tong Zhao}$^{\dag}$ \hspace{1.2em} \rm{Xupeng Miao}$^\dag$ \hspace{1.2em}  \rm{Jie Jiang}$^\S$ 
 \hspace{1.2em} Fangcheng Fu$^\ddag$ \hspace{1.2em} \rm{Bin Cui}$^\dag$
 \\ [.6em] 
 $^\dag$Peking University \hspace{1.2em} $^\ddag$Shanghai Jiao Tong University \hspace{1.2em} $^\S$Tencent \\ [.6em] 
}

\begin{abstract}
Mixture-of-Experts (MoE) and reinforcement learning (RL) post-training now dominate large language model (LLM) development, yet expert load imbalance remains a critical challenge. Existing load-balancing systems target pre-training by relying on historical step-level statistics. However, these methods fail under the unique workload dynamics of RL post-training: the step-level load is stable, but the tiny batch sizes processed during micro-steps cause severe, high-frequency load fluctuations.

We introduce \system, a micro-step-level load balancing system for MoE RL post-training. Instead of relying on historical statistics, \system exploits the multi-stage RL pipeline (rollout, recompute, policy update) by using foreseeable routing information from the rollout stage to proactively guide load balancing in the remaining stages. To support frequent per-micro-step reconfiguration, \system employs a hierarchical planner that decomposes the NP-hard load balancing problem into tractable sub-components, alongside a transfer engine that leverages complementary hardware paths (CPU-assisted and GPU-direct) for overlapped expert transfer. Evaluations on 64 GPUs demonstrate that \system achieves up to a 1.45$\times$ speedup over state-of-the-art RL post-training systems.
\end{abstract}

\maketitle
\blfootnote{*Equal contribution.}
\blfootnote{Contact: Yuming Zhou (ymzhou@stu.pku.edu.cn), Haoyang Li (lihaoyang@stu.pku.edu.cn), Fangcheng Fu (ccchengff@sjtu.edu.cn) and Bin Cui (bin.cui@pku.edu.cn).}


\section{Introduction}
\label{sec:intro}

The Mixture-of-Experts (MoE)~\cite{gshard} architecture has emerged as the dominant paradigm for scaling large language models (LLMs), powering recent models such as the DeepSeek~\cite{deepseek_v3, deepseek_r1}, Qwen~\cite{qwen_2, qwen_3}, and Kimi~\cite{kimi_k2, kimi_k2_5} series. 
Concurrently, reinforcement learning (RL) post-training, using algorithms like PPO~\cite{ppo}, GRPO~\cite{grpo}, and DAPO~\cite{dapo}, has proven essential for aligning LLMs with human preferences and unlocking advanced reasoning capabilities. As RL post-training is increasingly applied to massive MoE models (e.g., DeepSeek-R1~\cite{deepseek_r1}, Kimi K2.5~\cite{kimi_k2_5}), optimizing the efficiency of these training systems has become a critical challenge.

\begin{figure}[t]
\centering
\includegraphics[width=\columnwidth]{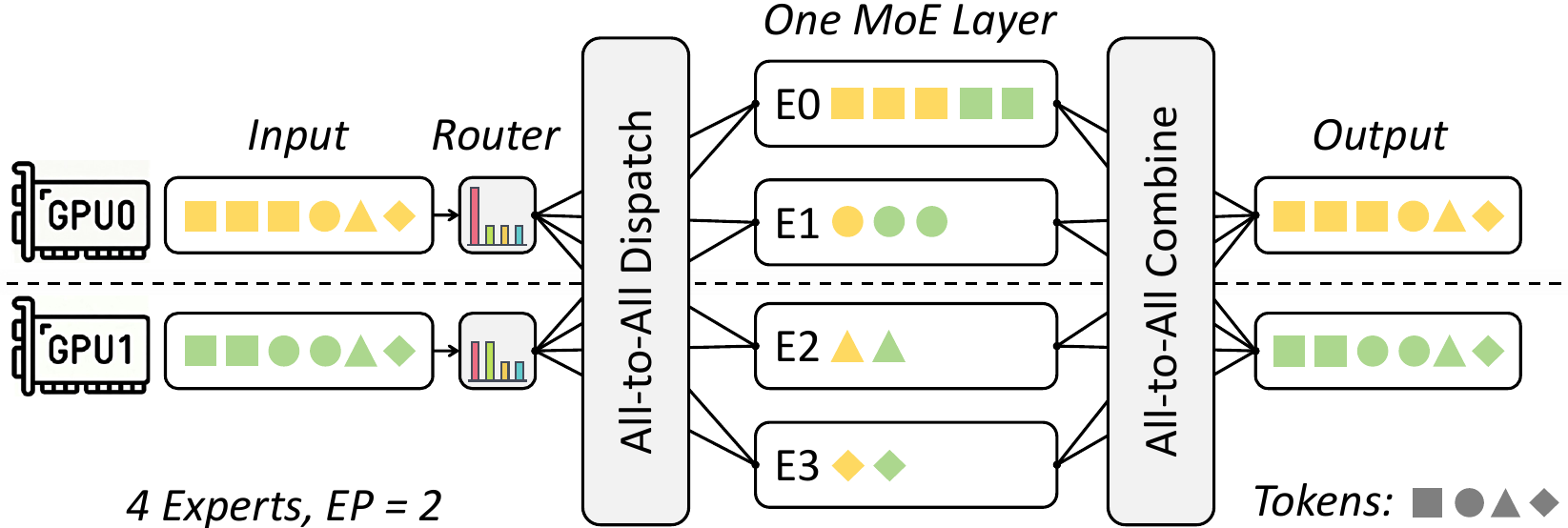}
\myvspace{-15pt}
\caption{\small{MoE layer with expert parallelism (EP). Load imbalance arises because different tokens are routed to different experts.}}
\label{fig:moe}
\myvspace{-5pt}
\end{figure}

However, as illustrated in Figure~\ref{fig:moe}, MoE training inherently suffers from load imbalance. The routing mechanism often assigns a disproportionate number of tokens to a few popular experts, leaving others underutilized. This significantly degrades overall training performance.

\begin{figure}[t]
\centering
\includegraphics[width=\columnwidth]{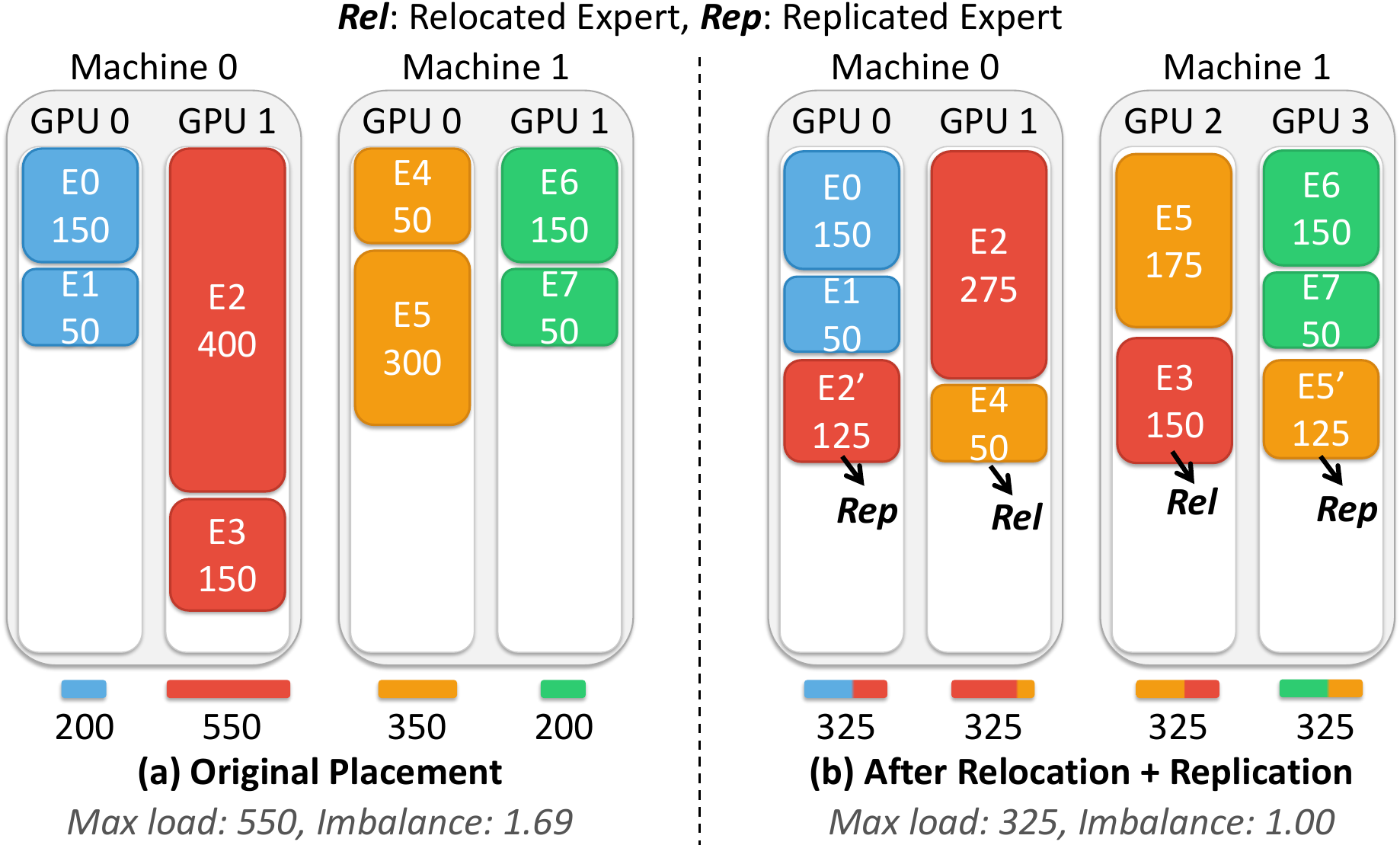}
\myvspace{-15pt}
\caption{\small{Expert relocation and replication mechanisms. (a) Original placement with severe load imbalance. (b) After jointly applying expert relocation (e.g., E3 and E4) and expert replication (e.g., E2' and E5'), perfect load balance is achieved.}}
\label{fig:mechanism}
\myvspace{-5pt}
\end{figure}

\begin{figure*}[t]
\centering
\includegraphics[width=\linewidth]{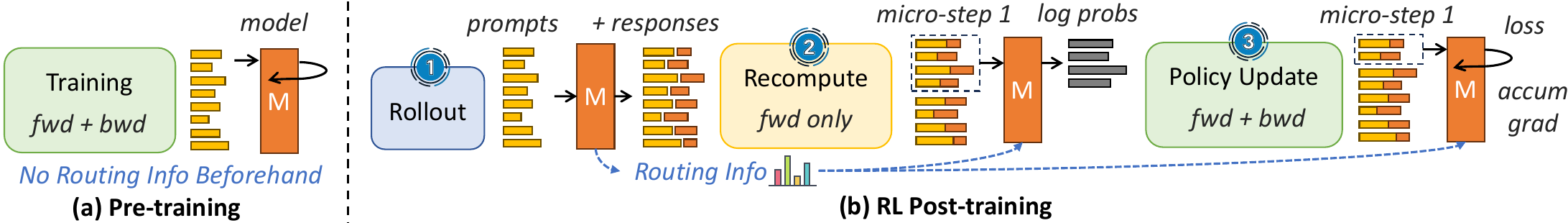}
\myvspace{-20pt}
\caption{\small{Comparison of (a) a pre-training step and (b) an RL post-training step for MoE models.
(a) During MoE pre-training, routing information is unavailable before execution, requiring existing approaches to predict routing behavior from historical statistics.
(b) An RL post-training step consists of rollout, recompute, and policy update stages. The rollout stage generates responses and records routing information, which is subsequently reused by the recompute and policy update stages, making the routing behavior foreseeable.}}
\label{fig:pipeline}
\end{figure*}

Prior efforts to alleviate expert load imbalance largely center on the pre-training phase~\cite{fastermoe, smartmoe, flexmoe, symi, eplb, popfetcher, laer_moe, fine_moe}. As Figure~\ref{fig:mechanism} shows, the primary mechanisms use expert relocation and replication to reconfigure expert placement, subsequently reassigning tokens to balance the load. However, effective reconfiguration requires precise knowledge of load patterns. Since routing behavior cannot be known in advance during pre-training (Figure~\ref{fig:pipeline}(a)), existing systems typically estimate the current load based on historical statistics from previous steps~\cite{fastermoe, smartmoe, flexmoe, symi, popfetcher, laer_moe}.

\mysubsubsection{Motivation: micro-step-level load balancing in RL.}
While these historical prediction methods work well for pre-training, we find they fall short in RL post-training. As shown in Figure~\ref{fig:pipeline}(b), RL post-training operates in three sequential stages, including rollout, recompute, and policy update. As detailed in \S\ref{sec:obs_and_mot}, RL post-training targets concentrated task domains (e.g., mathematics or coding) with model experts that have already specialized during pre-training. Consequently, the expert routing pattern converges and remains highly stable across steps. However, a stark instability emerges at the micro-step level: The recompute and policy update stages process data in numerous micro-steps (Figure~\ref{fig:pipeline}(b)), each containing few samples. This relatively small batch size per micro-step destroys the statistical averaging effect seen at the step level, causing severe, high-frequency fluctuations in expert load. Consequently, load balancing in RL post-training is inherently a micro-step-level problem.  Pre-training systems relying on step-level historical statistics simply cannot keep pace with these fine-grained variations. Furthermore, the latency budget available for planning and the possibility to amortize expert transfer costs largely vanish when reconfiguration must be performed at micro-step granularity.

\mysubsubsection{Challenges.}
Therefore, achieving effective \emph{micro-step-level load balancing} in RL post-training requires addressing three fundamental challenges: \textbf{(C1)} What signal should guide load balancing, if historical step-level statistics fail? \textbf{(C2)} How can we design an algorithm capable of solving the complex load-balancing problem at high, per-micro-step frequencies? \textbf{(C3)} How can we architect a system that supports the rapid expert transfers necessary for such frequent reconfigurations?

\mysubsubsection{Our Solutions.}
Fortunately, the characteristics of the RL post-training provide opportunities to address each of these challenges. To address \textbf{(C1)}, we recognize that the routing decisions made during the rollout stage are directly reused in the subsequent recompute and policy update stages (Figure~\ref{fig:pipeline}(b))~\cite{router_replay, stable_rl}. This structural trait allows us to use rollout routing information as a precise, foreseeable signal for proactive load balancing. 
To address \textbf{(C2)}, we leverage the observation that expert load is stable across steps. We decompose the original NP-hard load balancing problem into two parts: computing a stable base placement at the step level, and performing complementary adjustments at the micro-step level. This drastically reduces algorithmic overhead without sacrificing solving quality. Finally, to address \textbf{(C3)}, we observe that different RL stages naturally favor different expert transfer paths. CPU-assisted transfers over PCIe are optimal for the recompute stage, while GPU-direct transfers are more practical for the policy update stage (detailed in \S\ref{subsec:cpu_gpu_transfer}). Exploiting these complementary paths enables rapid expert transfers required for micro-step reconfigurations.

To this end, we introduce \system, a novel system that tackles the underexplored challenge of load imbalance in MoE RL post-training by enabling micro-step-level reconfiguration. \system harnesses foreseeable routing information from the rollout stage to orchestrate precise load balancing. At its core, \system features a \emph{Four-Stage Planner} that breaks the NP-hard load balancing problem into tractable sub-components, and an \emph{Expert Transfer Engine} that achieves fully overlapped reconfiguration by exploiting complementary transfer paths for different RL stages. 

In summary, our paper makes the following contributions:

\begin{itemize}[noitemsep, topsep=0pt, parsep=0pt, partopsep=0pt, leftmargin=*]
\item Our paper presents the first study of load imbalance characteristics in MoE RL post-training. Unlike pre-training, RL post-training exhibits highly stable step-level load distributions but severe micro-step-level fluctuations, highlighting the need for micro-step-level load balancing. 

\item We propose \system, which leverages foreseeable routing information for proactive micro-step load balancing. \system introduces algorithmic innovations to decompose the NP-hard load balancing problem hierarchically, alongside systemic innovations that utilize complementary hardware paths to achieve rapid expert transfer. 

\item Evaluations on MoE models on 64 GPUs demonstrate that \system achieves up to a 1.45$\times$ end-to-end speedup compared to state-of-the-art RL post-training systems.
\end{itemize}

\section{Preliminaries}
\label{sec:background}

In this section, we provide background on the MoE architecture, examine its load balancing issues, and present the RL post-training pipeline for MoE models.

\subsection{MoE Architecture}

The MoE~\cite{moe_survey_1, moe_survey_2, gshard} architecture replaces the standard feed-forward network (FFN) in each Transformer layer with a set of $E$ parallel expert networks and a lightweight router module. As shown in Figure~\ref{fig:moe}, for each input token, the router produces a probability distribution over all experts and selects the top-$K$ experts to process that token. The final output is the weighted sum of the selected experts' outputs, where the weights are the corresponding router probabilities. 

As shown in Figure~\ref{fig:moe}, Expert Parallelism (EP)~\cite{gshard} is used to partition experts across devices. Processing a MoE layer involves two phases of All-to-All collective communication: a dispatch phase that routes each token from its source rank to the rank hosting its assigned expert, and a combine phase that returns the expert outputs back to the source ranks. 

\subsection{MoE Load Balancing}




As depicted in Figure~\ref{fig:moe}, expert routing is often highly skewed, with a small subset of experts receiving a disproportionate number of tokens, leading to severe load imbalance across ranks. To address this issue, many prior works~\cite{fastermoe, smartmoe, flexmoe, symi, eplb, popfetcher, laer_moe} have studied load balancing for MoE pre-training. As shown in Figure~\ref{fig:mechanism}, their approaches can be broadly categorized into two mechanisms: expert relocation, which adjusts the expert-to-rank mapping, and expert replication, which places additional copies of heavily loaded experts.

Both expert relocation and expert replication rely on accurate estimates of expert loads. However, as illustrated in Figure~\ref{fig:pipeline}(a), routing information is unavailable before a pre-training step begins executing. Consequently, existing approaches predict the current step's expert loads from historical statistics observed in previous steps and use the resulting estimates to guide expert relocation and replication. 

While expert load balancing has been well-studied for pre-training, it remains largely underexplored in RL post-training. RL workloads exhibit distinct load dynamics that require  re-balancing at a much finer timescale, making pre-training methods fall short. We will elaborate on this in \S\ref{sec:obs_and_mot}. 

\subsection{RL Post-Training for MoE Models}

As shown in Figure~\ref{fig:pipeline}(b), RL post-training  typically follows three sequential stages:

\mysubsubsection{Rollout.} The model (i.e., the current policy) generates responses for a batch of prompts through auto-regressive decoding. This stage typically uses a high-throughput inference framework like vLLM~\cite{vllm} or SGLang~\cite{sglang}.

\mysubsubsection{Recompute.}
To address the inherent mismatch between rollout and training~\cite{stable_rl, flash_rl, qerl}, 
systems recompute log probabilities via a forward pass on the training framework (e.g., FSDP~\cite{fsdp} or Megatron~\cite{megatron_1, megatron_2}). These log probabilities will be used to correct the importance sampling~\cite{tis, mis, icepop} and become part of the loss in the subsequent policy update stage. Due to memory limits, sequences are split into micro-batches and processed sequentially in micro-steps.

\mysubsubsection{Policy Update.} The model is then optimized using a policy gradient objective (e.g., GRPO~\cite{grpo}). For memory efficiency, this stage also operates over sequential micro-steps to accumulate gradients before applying the final model update.

Throughout these three stages, routing consistency must be preserved (i.e., router replay)~\cite{stable_rl, router_replay}. Consequently, the routing information generated during rollout is recorded and reused throughout the following stages. As a result, unlike pre-training, the routing information for recompute and policy update stages is foreseeable prior to their execution.

\section{Observations and Current Limitations}
\label{sec:obs_and_mot}

\begin{figure}[t]
\centering
\includegraphics[width=\columnwidth]{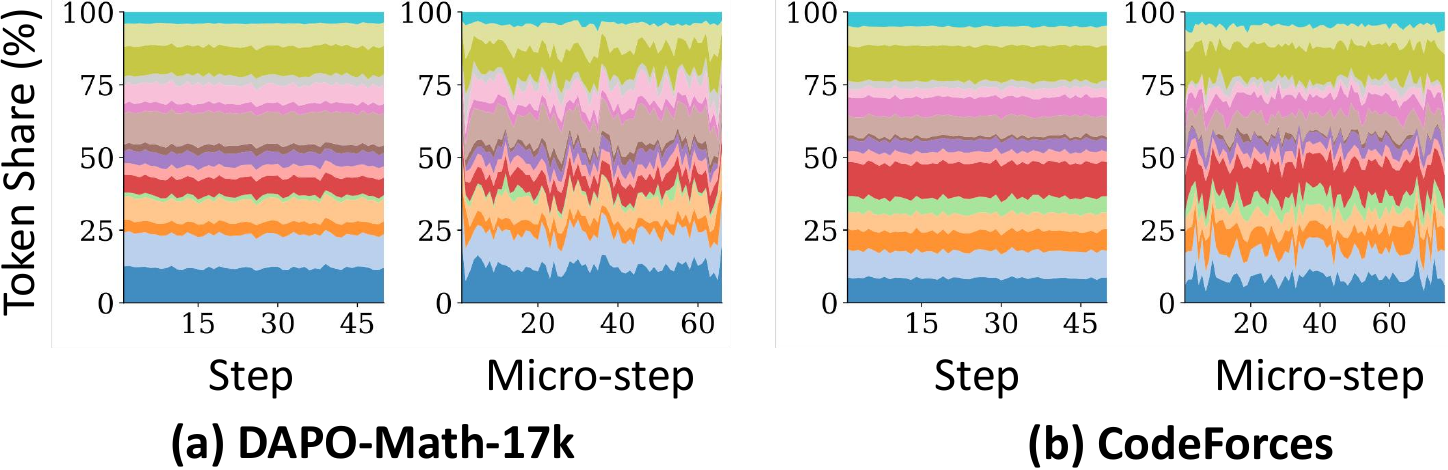}
\myvspace{-20pt}
\caption{\small{Expert load characteristics during RL post-training on Qwen3-30B-A3B. Step-level expert load remains stable but skewed, while micro-step-level expert load exhibits substantial fluctuations.}}
\label{fig:load_char}
\myvspace{-5pt}
\end{figure}

During pre-training, MoE models gradually develop expert specialization by learning from massive and diverse corpora, where different experts become specialized in distinct linguistic patterns~\cite{multilingual} or knowledge domains~\cite{myth_expert}. In contrast, RL post-training is typically performed on more concentrated tasks (e.g., mathematics~\cite{math_beyond, math_survey} or coding~\cite{swe_rl, coda_rl}) over a model whose experts have already been specialized during pre-training. Figure~\ref{fig:load_char} illustrates two distinctive characteristics of expert load during RL post-training on the DAPO-Math-17k~\cite{dapo_dataset} and CodeForces~\cite{codeforces_dataset} datasets, respectively.


\mysubsubsection{Step-level: stable but skewed.}
Because expert specialization has already been established before RL post-training, the overall routing distribution changes little across training steps. Consequently, when aggregated over a sufficiently large number of samples, expert loads exhibit only limited step-level variation. Meanwhile, the concentrated task domain tends to activate a subset of experts more frequently than others, causing some experts to consistently receive substantially more tokens. As a result, the long-term expert load distribution remains skewed.

\mysubsubsection{Micro-step-level: highly variable and skewed.}
Within a training step, however, each micro-step contains only a small number of samples whose routing behaviors may differ significantly. The reduced sample size weakens the statistical averaging effect observed at the step level, causing individual micro-steps to deviate substantially from the long-term load distribution. Consequently, expert loads can fluctuate dramatically across micro-steps, resulting in load distributions that are both highly variable and skewed.

\mysubsubsection{Current limitations.}
These observations suggest that load balancing in RL post-training should primarily target the micro-step timescale. However, existing MoE load-balancing techniques are predominantly designed for pre-training workloads~\cite{fastermoe, smartmoe, flexmoe, symi, eplb, popfetcher, fine_moe} 
and fall short for three reasons. First, they rely on historical step-level statistics and therefore cannot react to rapid load fluctuations at micro-step granularity. Second, re-balancing is typically triggered only after significant load shifts, allowing ample time for planning, whereas micro-step-level adaptation operates under much stricter latency budgets. Third, expert transfers are relatively infrequent in pre-training and their costs can be amortized across many steps, while micro-step-level reconfiguration requires substantially higher transfer efficiency.



\section{Challenges and Opportunities}
\label{sec:opp}

Therefore, achieving load balancing at the micro-step timescale in RL post-training requires addressing three key challenges:
\begin{itemize}[noitemsep, topsep=0pt, parsep=0pt, partopsep=0pt, leftmargin=*]
\item The step-level historical statistics widely used during pre-training are no longer informative. What information should guide load-balancing decisions instead?
\item From an algorithmic perspective, solving the optimal load-balancing plan is NP-hard (see \S\ref{sec:theory}), yet decisions must be made at micro-step timescales. How can we design the algorithm that achieves near-optimal load-balancing quality while meeting stringent real-time requirements?
\item From a systems perspective, micro-step-level balancing requires frequent reconfiguration throughout different micro-steps. How can experts be transferred efficiently enough to keep pace with such rapid reconfigurations?
\end{itemize}
To address these challenges, we identify three key opportunities, respectively, which together make micro-step-level load balancing both practical and effective.

\mysubsubsection{Opportunity 1: Exploiting foreseeable routing information as a planning signal.}
As shown in Figure~\ref{fig:pipeline}(b),  
by the time recompute or policy update begins, the routing behavior has already been determined during rollout and is fully observable. This exposes exact per-token, per-layer, and per-micro-step routing information, allowing the system to proactively plan load-balancing actions. 

\mysubsubsection{Opportunity 2: Exploiting the stability of step-level expert load.}
While expert load exhibits noticeable micro-step-level fluctuations, these fluctuations are not arbitrary: they remain centered around the aggregate step-level expert load distribution. This structure enables us to separate short-term load-balancing decisions from long-term ones.

Specifically, a stable expert placement can be derived to match the aggregate expert load distribution over training steps, while micro-step-level balancing focuses only on correcting localized deviations from this baseline. As a result, the full problem can be split into a base expert placement solving and a complementary adjustment per-micro-step. This decomposition preserves solving quality while substantially reducing the solving overhead. 

\mysubsubsection{Opportunity 3: Exploiting complementary expert transfer paths for different RL stages.}
Frequent micro-step load balancing requires equally frequent expert transfers. To make such reconfiguration practical, it is important to exploit efficient hardware paths. We identify two complementary transfer paths: (1) CPU-assisted transfer, where experts are fetched from CPU memory, and (2) GPU-direct transfer, which is widely adopted in prior work~\cite{fastermoe, smartmoe, flexmoe, symi, eplb}, where experts are obtained via GPU-to-GPU direct transfer. 
Importantly, we find that different RL stages impose fundamentally different requirements on expert transfers. The recompute stage is forward-only and therefore only involves transferring expert parameters. In contrast, the policy update stage includes both forward and backward computation, and must additionally handle gradient transfers and gradient accumulation across micro-steps. These differences fundamentally affect how expert transfers should be performed, creating an opportunity to employ the most suitable transfer path for each stage (detailed in \S\ref{subsec:cpu_gpu_transfer}).

\section{Overview of Our Solutions}
\label{sec:overview}

\begin{figure}[t]
\centering
\includegraphics[width=\columnwidth]{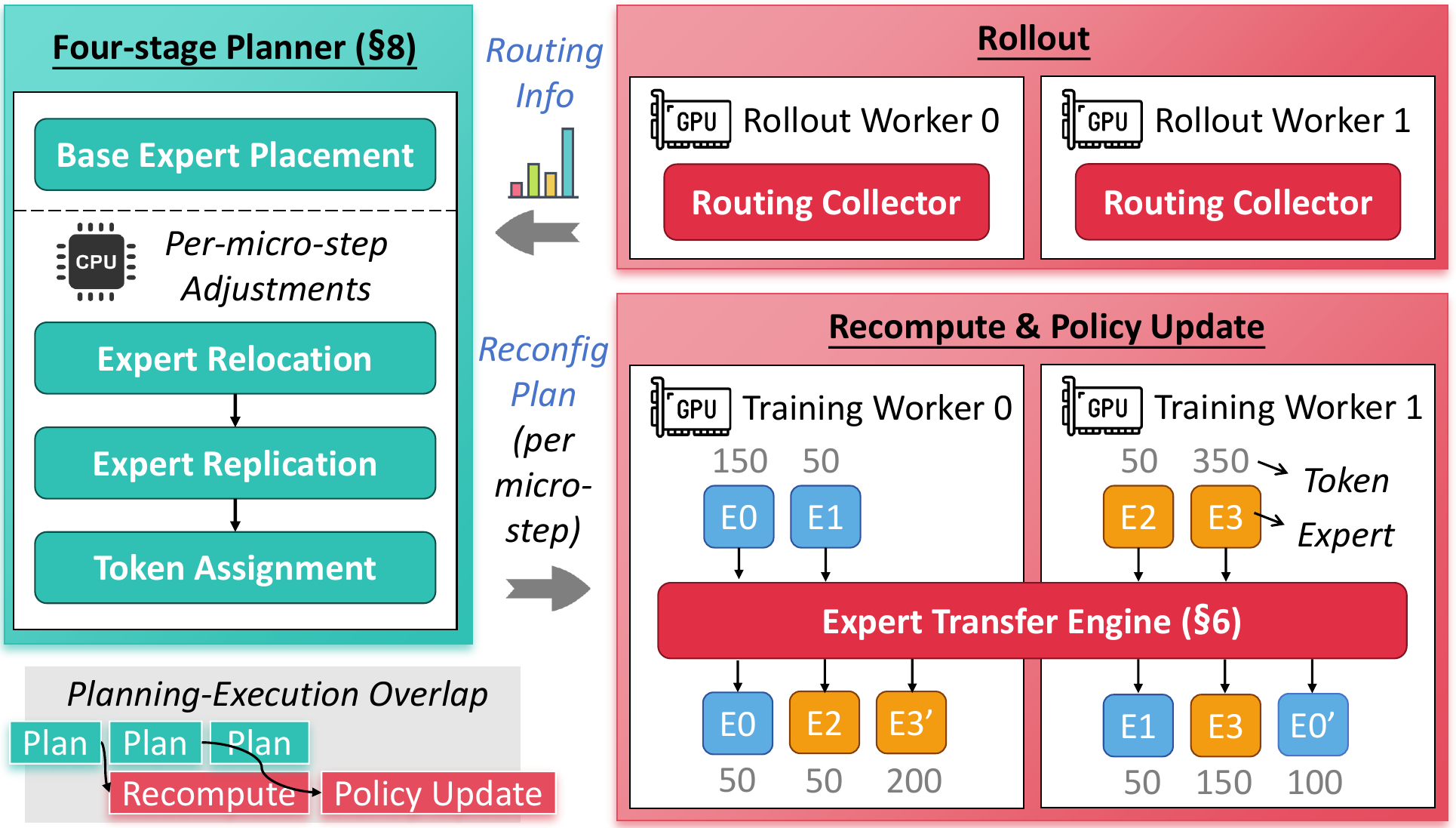}
\caption{\small{Overview of \system. The \emph{Rollout Collector} on each rollout worker collects routing information and feeds it to the \emph{Four-stage Planner} (\S\ref{sec:planner}). For each micro-step, the planner determines the optimal expert placement and token assignment. The \emph{Expert Transfer Engine} (\S\ref{sec:system}) then reconfigures expert placement as needed.}}
\label{fig:overview}
\end{figure}

Guided by the above opportunities, as shown in Figure~\ref{fig:overview}, we design \system, a system that leverages foreseeable routing  for micro-step-level MoE load balancing in RL post-training. 

To exploit routing information, \system introduces a \textit{Routing Collector} on each rollout worker. During the rollout stage, the collector records the router's top-$K$ expert selections for every token at every MoE layer, which provide the view of routing 
and serve as the foundation for planning.

To exploit the stability of step-level expert load, \system introduces a \textit{Four-stage Planner} that takes the routing information as input. The planner decomposes the original NP-hard optimization problem into a sequence of tractable subproblems and sequentially performs (1) base expert placement, (2) expert relocation, (3) expert replication, and (4) token assignment. While base expert placement is performed only once to match the aggregate expert load distribution across steps, the remaining three subproblems are solved for each micro-step to accommodate the micro-step-level load variations. This hierarchical planning process substantially reduces the solving overhead while preserving solution quality. Notably, it generates separate reconfiguration plans for the recompute and policy update stages, specifying expert placement and token assignment for each micro-step.

To exploit the complementary expert transfer paths for different RL stages, \system provides a unified \textit{Expert Transfer Engine} that supports both CPU-assisted and GPU-direct transfers. 
To minimize reconfiguration overhead, \system employs fine-grained communication--computation overlap to hide transfer costs, enabling efficient reconfiguration. 

As shown at the bottom of Figure~\ref{fig:overview}, throughout this workflow, the \textit{Four-stage Planner} executes on CPUs concurrently with GPU training. Furthermore, planning tasks for different layers and micro-steps are independent, enabling extensive parallelism. Consequently, the entire planning process can be fully overlapped with the training pipeline, incurring effectively zero critical-path overhead.

In the following, we first present the \emph{Expert Transfer Engine} (\S\ref{sec:system}) for efficient reconfiguration. We then introduce the theoretical formulation of the load-balancing problem (\S\ref{sec:theory}), followed by the \emph{Four-stage Planner} (\S\ref{sec:planner}), which decomposes the original NP-hard problem into tractable subproblems.
\section{Expert Transfer Engine}
\label{sec:system}

This section first presents two paths underlying the \textit{Expert Transfer Engine}: a CPU-assisted path and a GPU-direct path. We then discuss the rationale for using different transfer paths for different RL stages (\S\ref{subsec:cpu_gpu_transfer}). Building on these paths, we describe how they enable micro-step-level reconfiguration (\S\ref{subsec:per_micro_batch}) and further analyze the overhead (\S\ref{subsec:transfer_overhead}). 

\begin{figure}[t]
\centering
\includegraphics[width=\columnwidth]{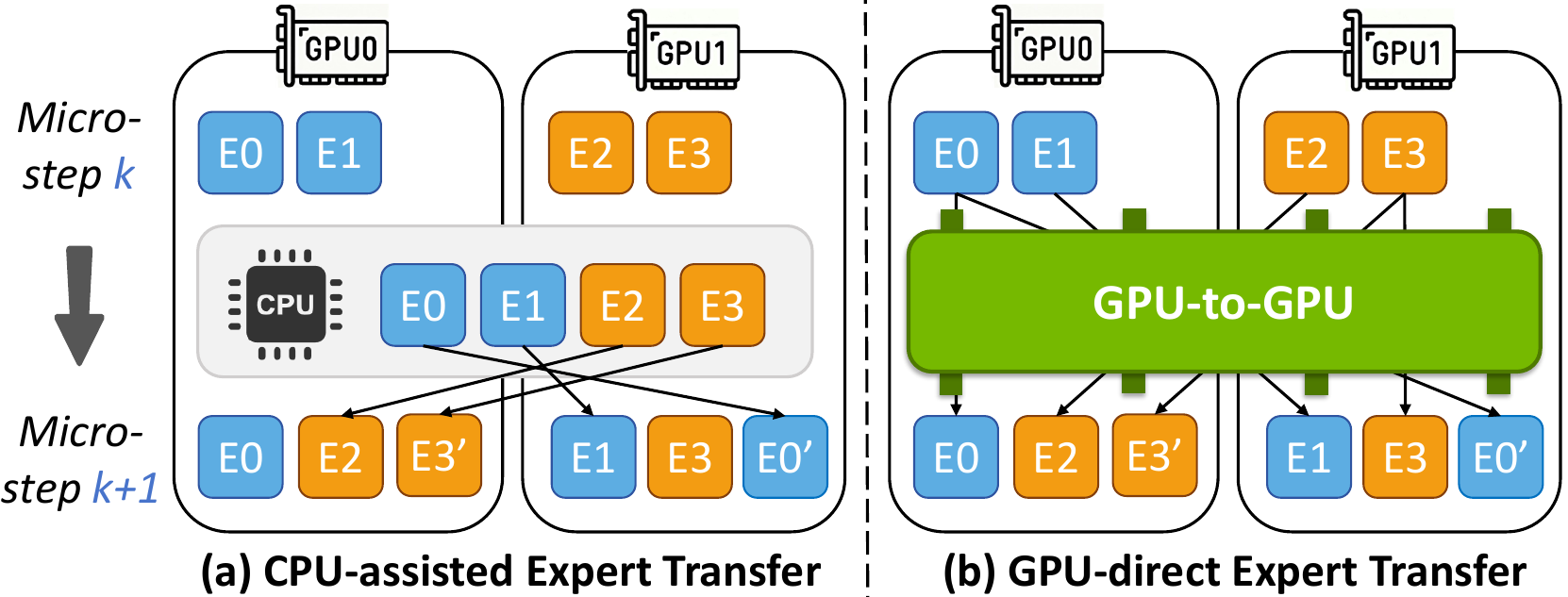}
\myvspace{-15pt}
\caption{\small{Two expert transfer paths.
(a) CPU-assisted path: each machine maintains a full copy of expert weights in pinned CPU memory. GPUs prefetch the experts required by the next micro-step via PCIe.
(b) GPU-direct path: expert weights reside on GPUs. Reconfiguration is performed through GPU-to-GPU transfers.
}}
\label{fig:cpu_gpu_system}
\myvspace{-5pt}
\end{figure}

\begin{figure*}[t]
\centering
\includegraphics[width=\linewidth]{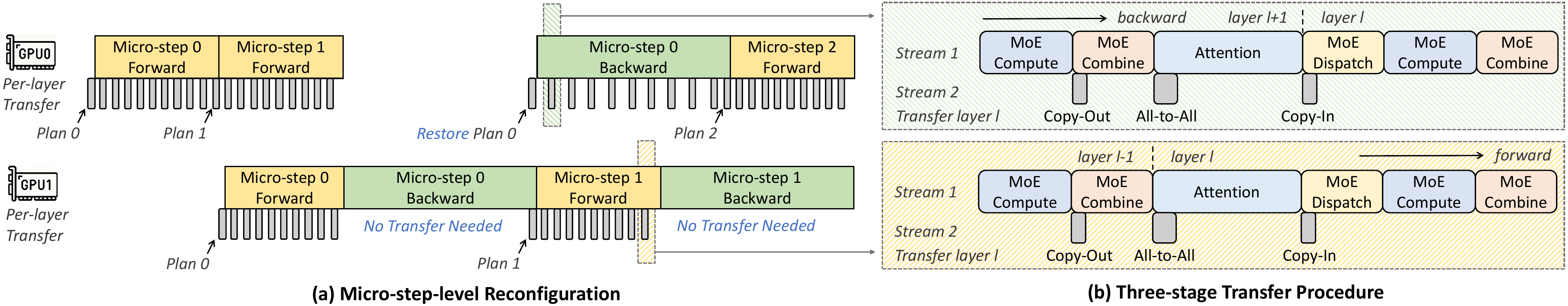}
\myvspace{-20pt}
\caption{\small{(a) Micro-step-level reconfiguration for the policy update stage, which includes both forward and backward passes. The forward and backward passes of the same micro-step use the same reconfiguration plan. (b) During both the forward and backward passes, per-layer expert transfer is overlapped with the execution on the main stream through a three-stage procedure.}}
\label{fig:overlap}
\end{figure*}

\subsection{Complementary Expert Transfer Paths}
\label{subsec:cpu_gpu_transfer}

\mysubsubsection{CPU-assisted expert transfer.}
Figure~\ref{fig:cpu_gpu_system}(a) illustrates the CPU-assisted path. Each machine keeps a full master copy of its layers' experts in pinned CPU memory, while only the experts currently in use occupy GPU memory. When the planner produces a per-micro-step plan, the engine reads which experts each rank needs for the upcoming micro-step, identifies those not already resident on the GPU, and prefetches them from CPU over PCIe. 

\mysubsubsection{GPU-direct expert transfer.} Figure~\ref{fig:cpu_gpu_system}(b) illustrates the GPU-direct path, which is widely adopted in prior MoE pre-training systems~\cite{fastermoe, smartmoe, flexmoe, symi, eplb}. 
Unlike prefetching from CPU over PCIe, relocation and replication are realized through GPU-to-GPU direct transfers. 

\mysubsubsection{Design rationale.}
In \system, we employ the CPU-assisted path during the recompute stage and the GPU-direct path during the policy update stage. The rationale behind this design choice is as follows.

Both transfer paths support communication--computation overlap by issuing expert transfers ahead of their actual use. However, their key difference lies in where the expert copies reside and which hardware channel carries the transfers. 
In the CPU-assisted design, the complete set of experts is maintained in CPU memory, allowing any GPU to easily fetch any expert. In contrast, the GPU-direct design can easily access only experts that already reside on GPUs within the same machine. Otherwise, if an expert resides on another machine, inter-machine transfers are triggered, which are substantially more expensive than intra-machine transfers over NVLink, making the overhead difficult to hide (as evaluated in \S\ref{subsec:eval_ablation}). Consequently, we restrict the GPU-direct path to be planned only within a machine. Under this restriction, CPU-assisted transfer is theoretically preferable because it provides unrestricted access to the full expert pool, thus exposing a much larger placement space to the planner.

The recompute stage therefore adopts the CPU-assisted path, as it only involves expert parameter transfers. The policy update stage, however, must additionally handle gradient transfers. Unlike parameters, gradients must be accumulated across micro-steps. Such frequent updates would require continuous synchronization between CPU and GPU if gradients were maintained on the CPU, introducing substantial overhead. We therefore adopt the GPU-direct path for the policy update stage. Further discussion and a feasibility analysis of using a CPU-assisted variant for the policy update stage are provided in Appendix~\ref{app:cpu_pol}.

\subsection{Micro-step-level Reconfiguration}
\label{subsec:per_micro_batch}

Prior MoE pre-training systems typically perform reconfiguration at the step level~\cite{fastermoe, smartmoe, flexmoe, symi}. The low frequency makes serial expert transfers practical, as the overhead is small relative to step time. Micro-step-level reconfiguration, by contrast, demands much more frequent transfers, so the overhead can no longer be easily amortized. Hiding this overhead thus becomes critical to efficient reconfiguration.


\mysubsubsection{Per-layer expert transfer.}
To hide this overhead, we design a per-layer transfer procedure. The key idea is to dedicate a separate CUDA stream to transfer the experts required by the next layer. By initiating the transfer ahead of time, communication on the auxiliary stream can overlap with the computation of the current layer on the main stream.

For the recompute stage, which consists only of forward pass and involves parameter transfers alone, the CPU-assisted path enables experts to be prefetched from CPU memory directly. As a result, expert transfers can be easily overlapped.

The policy update stage is more challenging. As illustrated in Figure~\ref{fig:overlap}(a), it consists of both forward and backward passes, requiring distinct overlap mechanisms for the two passes. Furthermore, expert transfer in this stage involves both parameters and gradients. To improve bandwidth utilization, we should aggregate them into a single transfer. Based on these considerations, we design the three-stage procedure shown in Figure~\ref{fig:overlap}(b):


\begin{itemize}
[leftmargin=*]
\item \textbf{Copy-out.} Pack the parameters and gradients that need to leave the current GPU 
into a contiguous send buffer.
\item \textbf{All-to-All swap.} Exchange packed buffers.
\item \textbf{Copy-in.} 
Unpack the parameters and gradients from the receive buffer into the target memory on the GPU. For the backward pass in particular, for each replicated expert, we designate one replica as the \emph{main expert}, and accumulate the partial gradients from all replicas into its gradient so that the optimizer applies a single update.
\end{itemize}

For both the forward and backward passes, the three stages of layer $l$'s transfer are scheduled against the surrounding training operations as shown in Figure~\ref{fig:overlap}(b): Copy-out overlaps with the MoE combine, All-to-All swap overlaps with the Attention, and Copy-in overlaps with the MoE dispatch. 

\mysubsubsection{Reconfiguration plan management.}
While the planner continues to generate reconfiguration plans, the \emph{Expert Transfer Engine} maintains the plans (for both recomputation and policy updates) for all unexecuted micro-steps. In the recompute stage, a plan is consumed and discarded immediately after the micro-batch’s forward pass. The policy update stage, however, requires the same plan to serve both the forward and backward passes of a micro-step. As shown in Figure~\ref{fig:overlap}(a), under pipeline schedules such as 1F1B~\cite{pipedream, pipedream_2bw}, the backward pass may execute long after the forward pass, when the GPU placement has already been reconfigured for other micro-steps. Thus, the plan must be retained until the backward pass completes, allowing it to restore the original forward-time placement by replaying the old plan.

\subsection{Overhead Analysis}
\label{subsec:transfer_overhead}

Though the reconfiguration happens per-micro-step, both CPU-assisted and GPU-direct transfer (within a machine) incur minimal overhead in terms of memory and latency. (1) Regarding memory, the CPU-assisted path introduces no additional GPU memory consumption, while the GPU-direct path requires only minimal GPU memory for temporary send/receive buffers. These buffers are released immediately after communication and therefore do not contribute to peak memory usage. (2) Regarding latency, empirical evaluations in \S\ref{subsec:eval_overhead} demonstrate that the overhead can be fully overlapped for both paths. Appendix~\ref{app:overhead} further derives the conditions for overlapping per-micro-step, per-layer expert transfers, showing that the required per-rank sequence length is only a few thousand tokens for both paths, well below that of typical RL post-training workloads. 


\section{Theoretical Modeling}
\label{sec:theory}

This section presents the theoretical formulation of our micro-step-level load balancing problem. We first develop a time model to estimate the execution time of a single MoE layer, capturing both computation and communication costs (\S\ref{subsec:moe_time}). Based on this model, we show that, given the routing information of a micro-step, jointly optimizing expert placement and token assignment can be formulated as a Mixed-Integer Linear Program (MILP) (\S\ref{subsec:milp}). 

\mysubsubsection{Terminology.}
Table~\ref{tab:notation} summarizes the notation used throughout the following sections. We elaborate on two key terms that appear repeatedly in our formulation:

\begin{itemize}[leftmargin=*]
\item \textbf{Rank.} An EP rank is a single device (i.e., GPU) within the EP group. There are $P$ ranks in total, distributed evenly across $M$ machines, with $P/M$ ranks per machine.
\item \textbf{Slot.} Each rank is configured with $N_s$ \emph{slots}, where every slot can hold the parameters as well as the gradients of one expert. Of the $N_s$ slots on a rank, $N_b = E/P$ are \emph{base slots} (enough to cover every expert in the model exactly once if there were no replication), and the remaining $N_r$ are \emph{redundant slots} available for replicated experts (i.e., $N_r = N_s - N_b$). Across the whole EP group there are $P \cdot N_s$ slots in total, indexed by $j \in \{1, 2, \ldots, P \cdot N_s\}$. 
\end{itemize}

\begin{table}[t]
\centering
\caption{\small{Notation used in our problem formulation.}}
\myvspace{-5pt}
\label{tab:notation}
\small
\begin{tabular}{cl}
\toprule
Symbol & Description \\
\midrule

$E, P, M$
& Number of experts, EP ranks, and machines \\

$N_b, N_r, N_s$
& Base, redundant and total slots
per rank \\

$w_{s,e}$
& Token volume from source rank $s$ to expert $e$ \\

$x_{e,j}$
& Binary, $1$ if expert $e$ is assigned to slot $j$ \\

$r_{s,e,j}$
& Fraction of $w_{s,e}$ routed to slot $j$ \\

$L_r$
& Total token load processed by rank $r$ \\

$C_{i,j}$
& Traffic volume from machine $i$ to machine $j$ \\

$ \texttt{slots}(r \mid m)$
& Slots owned by rank $r$ or machine $m$ \\

$ \texttt{ranks}(m)$
& Ranks owned by machine $m$ \\

$\texttt{machine}(r)$
& Machine owning rank $r$ \\

\bottomrule
\end{tabular}
\myvspace{-5pt}
\end{table}

\subsection{MoE Layer Time Model}
\label{subsec:moe_time}

We model the execution time of a single MoE layer as the sum of computation and communication time. Both are bounded by the slowest rank, as All-to-All barriers force every rank to wait for the slowest one.

\mysubsubsection{Computation time.} 
The computation time is set by the rank with the largest token load:
\begin{equation}
\small
T_{\text{comp}} = K_1 \cdot L_{\max} + B_1,
\label{eq:t_comp}
\end{equation}
where $L_{\max} = \max_{r \in [P]} L_r$ is the maximum token load, $K_1$ is the coefficient, and $B_1$ is a fixed overhead per layer.

\mysubsubsection{Communication time.} 
Communication is dominated by the most heavily loaded inter-machine directional link. In contrast, intra-machine traffic traverses the much faster NVLink fabric and is therefore not a bottleneck. Accordingly, the communication time can be modeled as:
\begin{equation}
\small
T_{\text{comm}} = K_2 \cdot C_{\max} + B_2,
\label{eq:t_comm}
\end{equation}
where $C_{\max} = \max_{i \neq j} C_{i,j}$ is the maximum inter-machine traffic, $K_2$ is the coefficient, and $B_2$ is a fixed latency. 

\mysubsubsection{Total layer time.} Summing the two components,
\begin{equation}
\small
T_{\text{MoE}} = n_1 \cdot (K_1 \cdot L_{\max} + B_1) + n_2 \cdot (K_2 \cdot C_{\max} + B_2),
\label{eq:layer_time_full}
\end{equation}
where $n_1$ and $n_2$ are the numbers of computation and communication rounds invoked by one layer pass. For the recompute stage (one forward), $n_1 = 1$ and $n_2 = 2$ (one MoE dispatch and one MoE combine). For the policy update stage (forward plus backward), $n_1 = 3$ and $n_2 = 4$, since the backward pass triggers an additional gradient computation and additional MoE dispatch and MoE combine of gradients.

This time model isolates the two metrics the placement problem must minimize: the maximum rank load $L_{\max}$ and the maximum inter-machine traffic $C_{\max}$. Both depend on how experts are placed across slots and how tokens are assigned to those slots. 

\subsection{Joint MILP Formulation}
\label{subsec:milp}

We can formulate the micro-step-level expert placement and token assignment problem for minimizing the per-layer $T_{\text{MoE}}$ as a Mixed-Integer Linear Program (MILP). The inputs consist of the per-source, per-expert token volumes $w_{s,e}$ for $s \in [P]$ and $e \in [E]$, together with the constants $P$, $M$, $N_b$, $N_r$, and $N_s$. The decision variables are:

\begin{itemize}[leftmargin=*]
\item $x_{e,j} \in \{0,1\}$: \emph{placement variable}. $x_{e,j}=1$ if and only if expert $e$ is placed in slot $j$.
\item $r_{s,e,j} \in [0,1]$: \emph{assignment variable}. $r_{s,e,j}$ denotes the fraction of token volume $w_{s,e}$ that rank $s$ routes to slot $j$.
\end{itemize}

\mysubsubsection{Derived metrics.} The rank load $L_r$ aggregates all tokens that are assigned to any slot owned by rank $r$:
\begin{equation}
\small
L_r = \sum_{s\in[P]} \sum_{e\in[E]} \sum_{j \in \texttt{slots}(r)} w_{s,e} \cdot r_{s,e,j}.
\label{eq:rank_load}
\end{equation}
The inter-machine traffic $C_{i,j}$ aggregates all tokens that flow from  ranks on machine $i$ to slots owned by machine $j$: 
\begin{equation}
\small
C_{i,j} = \sum_{s \in \texttt{ranks}(\mathrm{machine}_i)} \sum_{e\in[E]} \sum_{j' \in \texttt{slots}(\mathrm{machine}_j)} w_{s,e} \cdot r_{s,e,j'},\; i \neq j.
\label{eq:rail_traffic}
\end{equation}

\mysubsubsection{Constraints.} A valid expert placement and token assignment must satisfy the following four constraints, corresponding to expert slot capacity, expert coverage, token conservation, and token assignment feasibility, respectively (see Appendix~\ref{app:milp_constraints} for a detailed explanation of each constraint):
{\small
\begin{align}
\textstyle\sum_{e=1}^{E} x_{e,j} = 1,& \quad \forall j \in [P \cdot N_s], 
\\
\textstyle\sum_{j=1}^{P \cdot N_s} x_{e,j} \geq 1, & \quad \forall e \in [E], 
\\
\textstyle\sum_{j=1}^{P \cdot N_s} r_{s,e,j} = 1, &\quad \forall s \in [P],\, e \in [E], 
\\
r_{s,e,j} \leq x_{e,j}, & \quad \forall s\in [P],\,e\in [E],\,j\in [P \cdot N_s]. 
\end{align}
}

\mysubsubsection{Objective.} The objective is to minimize $T_{\text{MoE}}$ as defined in Eq.~(\ref{eq:layer_time_full}). This objective is inherently nonlinear because $L_{\max}$ and $C_{\max}$ are formulated as max functions. To address this, we apply the standard epigraph trick~\cite{epigraph}, replacing each maximum with an auxiliary variable. This yields a proper Mixed-Integer Linear Program (MILP). 

\mysubsubsection{Complexity.} This MILP has $E \cdot P \cdot N_s$ binary variables $x_{e,j}$ and $E \cdot P^2 \cdot N_s$ continuous variables $r_{s,e,j}$, coupled by the constraint $r_{s,e,j} \leq x_{e,j}$. It is NP-hard in general, and must be solved for every (micro-step, layer) pair. Solving it within the planner's per-micro-step time budget is therefore infeasible. 

\section{Four-stage Planner}
\label{sec:planner}

The complexity of the original MILP problem motivates the design of our \textit{Four-stage Planner}, which decomposes the per-(micro-step, layer) MILP into four sequential stages: base expert placement (Stage 1), expert relocation (Stage 2), expert replication (Stage 3) and token assignment (Stage 4).

This four-stage decomposition directly accommodates the temporal patterns identified in \S\ref{sec:obs_and_mot}. Specifically, Stage 1 captures the step-level stability: since the per-step aggregate load remains nearly constant across consecutive training steps, a single base mapping computed from this aggregate is reusable over many steps (\S\ref{subsec:base_placement}). In contrast, Stages 2–4 absorb the micro-step-level volatility: each micro-step's empirical load deviates from the aggregate, and these stages adjust both expert placement and token assignment to fit the per-micro-step load variation (\S\ref{subsec:per_micro_step_adjust}).


\subsection{Stage 1: Base Expert Placement}
\label{subsec:base_placement}

Stage 1 assigns every expert to a single base slot using only the per-step aggregate load $\bar w_{s,e} = \sum_i w_{s,e}^{(i)}$, where $w_{s,e}^{(i)}$ denotes the load of expert $e$ from source rank $s$ in the $i$-th micro-step. The output $A_{\text{base}}$ is a placement that fills the $N_b$ base slots on every rank. The remaining $N_r$ redundant slots per rank remain empty and are filled during Stage 3.

We adopt the hierarchical greedy procedure shown in Algorithm~\ref{alg:base}. It first determines a machine-level placement that specifies the hosting machine of each expert, and then refines it into a rank-level placement that distributes each machine's assigned experts across its local ranks.

\begin{algorithm}[t]
\caption{\small{Stage 1: Base Expert Placement}}
\label{alg:base}
\small
\begin{algorithmic}[1]
\Require Experts $E$, EP size $P$, machines $M$,  Per-step aggregate load matrix $\bar W$ of size $P \times E$
\Ensure Base expert placement $A_{\text{base}} : [E] \to [P]$
\State $\bar w_e \gets \sum_s \bar w_{s,e}$ for all $e$
\State Sort experts in descending order of $\bar w_e$
\State $\text{ML}[m] \gets 0$, $\text{MC}[m] \gets 0$ for $m \in [M]$
\State \textbf{// Machine-level placement}
\For{each expert $e$ in sorted order}
    \For{$m = 1$ to $M$}
        \State $\Delta_{m,e} \gets \sum_{s : \texttt{machine}(s) \neq m} \bar w_{s,e}$
        \State $\texttt{score}(m,e) \gets n_1 K_1 (\text{ML}[m] + \bar w_e) + n_2 K_2 (\text{MC}[m] + \Delta_{m,e})$
    \EndFor
    \State $m^* \gets \arg\min_m \texttt{score}(m, e)$, place expert $e$ in machine $m^*$
    \State $\text{ML}[m^*] \mathrel{+}= \bar w_e$; $\text{MC}[m^*] \mathrel{+}= \Delta_{m^*,e}$
\EndFor
\State \textbf{// Rank-level placement}
\For{each machine $m$}
    \State $\text{RL}[r] \gets 0$ for $r \in \texttt{ranks}(m)$
    \State Sort experts on $m$ by descending $\bar w_e$
    \For{each expert $e$ on $m$}
        \State $r^* \gets \arg\min_{r \in \texttt{ranks}(m)} \text{RL}[r]$
        \State $A_{\text{base}}[e] \gets r^*$; $\text{RL}[r^*] \mathrel{+}= \bar w_e$
    \EndFor
\EndFor
\State \Return $A_{\text{base}}$
\end{algorithmic}
\end{algorithm}

\mysubsubsection{Machine-level placement}. Experts are sorted by descending aggregate load $\bar w_e = \sum_s \bar w_{s,e}$ and placed greedily. For each expert, we select the machine that minimizes a joint cost combining computation load and cross-machine traffic. In Algorithm~\ref{alg:base}, $\mathrm{ML}[m]$ tracks machine $m$'s accumulated computation load, and $\mathrm{MC}[m]$ tracks its accumulated inbound cross-machine traffic. Placing expert $e$ on machine $m$ would raise $\mathrm{ML}[m]$ by $\bar w_e$ and raise $\mathrm{MC}[m]$ by $\Delta_{m,e} = \sum_{s\,:\,\texttt{machine}(s) \neq m} \bar w_{s,e}$,
the volume of tokens arriving from other machines to reach $e$. Weighting the two increments by the time-model coefficients of \S\ref{subsec:moe_time} gives the scoring function $\texttt{score}(m, e)$. 
We place $e$ on machine $m^* = \arg\min_m \texttt{score}(m, e)$ and update $\mathrm{ML}[m^*]$, $\mathrm{MC}[m^*]$  accordingly. 

\mysubsubsection{Rank-level placement.} Within each machine, we then distribute its placed experts across the local ranks using the Longest Processing Time (LPT)~\cite{lpt} heuristic: In Algorithm~\ref{alg:base}, $\mathrm{RL}[r]$ tracks rank $r$'s accumulated computation load. Experts are processed in descending order of load and each is placed on the currently least-loaded local rank. This step balances per-rank computation load within each machine without affecting any cross-machine traffic decided during the machine-level placement.

\subsection{Stages 2--4: Per-micro-step Adjustments}
\label{subsec:per_micro_step_adjust}


Once the base expert placement $A_\text{base}$ is determined, each micro-step only needs to adjust the expert placement and determine the token assignment around this base. Algorithm~\ref{alg:per_micro_step} outlines this procedure.

\begin{algorithm}[t]
\caption{\small{Stages 2-4: Per-micro-step Adjustments}}
\label{alg:per_micro_step}
\small
\begin{algorithmic}[1]
\Require Base expert placement $A_{\text{base}}$, per-micro-step loads $\{W^{(i)}\}_{i=1..N}$, redundant slots per rank $N_r$, max relocation rounds $T$
\Ensure Per-micro-step expert placements $\{A^{(i)}\}$ and token assignments $\{r^{(i)}\}$
\For{each micro-step $i = 1..N$ \textbf{in parallel}}
    \State $A^{(i)} \gets A_{\text{base}}$
    \State Init loads $\text{RL}[r]{=}L_r$ and traffic $\text{LT}[i,j]{=}C_{i,j}$ from $A^{(i)}, W^{(i)}$
    \State \textbf{// Stage 2: expert relocation via swaps}
    \For{$round = 1$ to $T$}
        \State $h \gets \arg\max_r \text{RL}[r]$ \Comment{bottleneck rank}
        \State For each $r_l \neq h$, evaluate swapping a top-$K$ heaviest expert on $h$ with a top-$K$ lightest expert on $r_l$
        \State $(e_h, e_l, r_l)^\star \gets$ swap minimizing $\Delta$ \Comment{$\Delta$: change in layer-time objective}
        \If{$\Delta \geq 0$} \textbf{break} \EndIf
        \State Apply swap; update $\text{RL}, \text{LT}$
    \EndFor
    \State \textbf{// Stage 3: expert replication via redundant slots}
    \For{$k = 1$ to $P \cdot N_r$}
        \State $(e^\star, r^\star, \Delta) \gets$ replica candidate with largest estimated drop (locality-aware assignment)
        \If{$\Delta \geq 0$} \textbf{break} \EndIf
        \State Place a replica of $e^\star$ on $r^\star$
    \EndFor
    \State \textbf{// Stage 4: token assignment via LP}
    \State Solve the LP with $\{x_{e,j}\}$ fixed by Stages 2-3 for $\{r^{(i)}\}$
\EndFor
\State \Return $\{A^{(i)}\}, \{r^{(i)}\}$
\end{algorithmic}
\end{algorithm}

\mysubsubsection{Stage 2: expert relocation.} The base placement is no longer optimal because each micro-step's load matrix $w^{(i)}_{s,e}$ deviates from the aggregate $\bar w_{s,e}$. To correct this, we first leverage expert relocation by repeatedly swapping pairs of experts. At each round, the algorithm selects the most heavily loaded rank as the swap source, pairs it against every other rank as a swap target, and evaluates a small top-$K \times $ top-$K$ window of candidate expert pairs per target. The swap that yields the largest reduction in the layer time objective is committed. Restricting the source to the bottleneck rank and bounding the per-target candidate set by $K$ keeps each round's cost at $O(P \cdot K^2)$. 
This loop terminates when no swap improves the objective or the maximum round count is reached.

\mysubsubsection{Stage 3: expert replication.} With Stage 2's placement fixed, $P \cdot N_r$ redundant slots remain. We allocate them one at a time. At each step, for every (expert, rank) candidate, we estimate the objective reduction obtained by adding that replica using a locality-aware token assignment heuristic: tokens prefer same-machine replicas to avoid cross-machine traffic, and any leftover tokens are routed to minimize per-rank load. The candidate yielding the largest reduction is selected. The loop terminates when all redundant slots are filled or no candidate improves the objective.

\mysubsubsection{Stage 4: token assignment.} Once the expert placement is all fixed, the binary variables $x_{e,j}$ become constants and the MILP of \S\ref{subsec:milp} collapses to a linear program (LP) over the continuous token assignment variables $r_{s,e,j}$. Concretely, given the placement $\{x_{e,j}\}$ produced by Stages 2-3, the LP for one (micro-step, layer) instance is
\begin{equation}
\small
\begin{aligned}
\min_{\,r,\,L^*,\,C^*} \quad & n_1 K_1 L^* + n_2 K_2 C^* \\
\text{s.t.} \quad
& \textstyle\sum_{j} r_{s,e,j} = 1,
  && \forall s \in [P],\, e \in [E], \\
& 0 \le r_{s,e,j} \le x_{e,j},
  && \forall s, e, j, \\
& L_r \le L^*,
  && \forall r \in [P], \\
& C_{i,j} \le C^*,
  && \forall i \neq j,
\end{aligned}
\label{eq:lp_token}
\end{equation}
where $L_r$ and $C_{i,j}$ are the rank load and cross-machine traffic defined in \S\ref{subsec:milp}. We solve Eq.~(\ref{eq:lp_token}) with the HiGHS~\cite{highs} solver to obtain optimal token assignment. Three implementation optimizations keep the per-micro-step LP cheap: (1) only replicated experts contribute non-trivial decision variables, since experts with a single placement have a deterministic token assignment and can be eliminated, (2) the constraint matrix is constructed in sparse form via vectorized operations, and (3) because (micro-step, layer) instances are independent, LPs are solved in parallel across CPU cores.

Algorithm~\ref{alg:per_micro_step} forms the basis for both the recompute stage and the policy update stage. In the policy update stage, however, expert relocation (Stage 2) and expert replication (Stage 3) are subject to certain restrictions: both swapping and replica addition must consider whether the reconfiguration can be performed via the GPU-direct path within a machine. By contrast, the recompute stage faces no such limitations, as it can draw on the entire expert pool for relocation and replication through the CPU-assisted transfer path. We refer interested readers to Appendix~\ref{app:update_alg} for the adapted algorithm under the restrictions during the policy update stage.

\section{Implementation}
\label{sec:impl}

We implement \system on top of veRL~\cite{verl}, with approximately 10K lines of Python code. We extend Megatron~\cite{megatron_1, megatron_2} to support dynamic expert placement for training and use vLLM~\cite{vllm} for rollout.

\mysubsubsection{Planner.} The \emph{Four-stage Planner} is implemented as a standalone Ray~\cite{ray} actor. It leverages CPU cores across the entire cluster to plan different (micro-step, layer) instances in parallel, enabling scalable decision-making. As a result, planning is fully decoupled from GPU training and can be completely overlapped with the training process. 



\mysubsubsection{Token assignment across replicated experts.} Standard EP implementations dispatch each token to exactly one destination rank. 
With expert replication, a token may have multiple dispatch options. We implement a custom token dispatch kernel in Triton~\cite{triton} that extends the standard All-to-All dispatch to support further token assignment among replicas. This Triton implementation avoids modifying the underlying NCCL~\cite{nccl, nccl_doc} All-to-All and maintains compatibility with existing EP communication infrastructure.

\begin{table}[t]
\centering
\caption{\small{Models, parallelism configurations, and GPUs used for training (the remainder are used for rollout).} DP, TP, and EP denote the data-, tensor-, and expert-parallelism degrees.}
\myvspace{-5pt}
\label{tab:model_config}
\small
\begin{tabular}{lccccc}
\toprule
Model & \# Experts & $\left<DP, TP, EP\right>$ & \# Train GPUs \\
\midrule
Qwen3-30B-A3B   & 128 & $\left<4, 4, 16\right>$ & 16 (Total: 64)\\
Qwen3-30B-A3B   & 128 & $\left<8, 4, 32\right>$ & 32 (Total: 64) \\
Qwen3.5-35B-A3B & 256 & $\left<16, 2, 32\right>$ & 32 (Total: 64) \\
\bottomrule
\end{tabular}
\myvspace{-5pt}
\end{table}

\begin{figure*}[t]
\centering
\includegraphics[width=\linewidth]{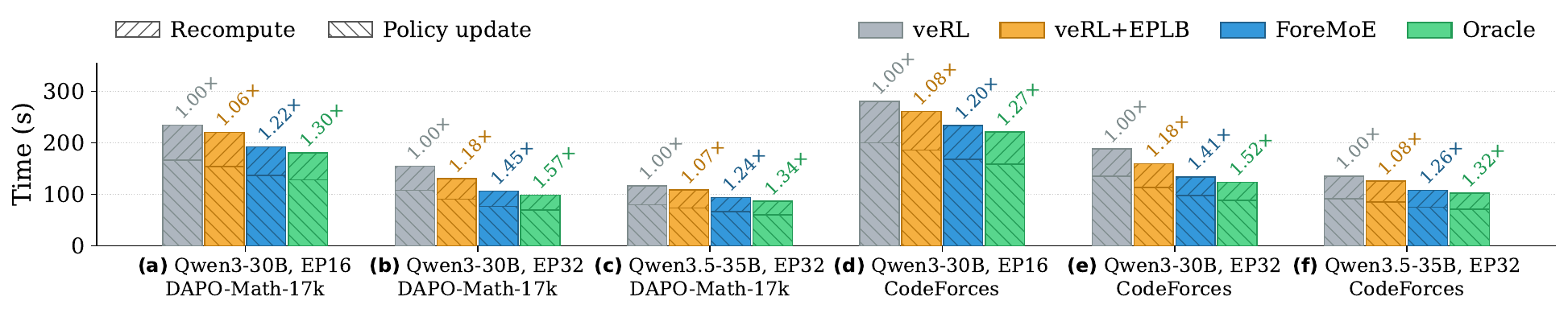}
\myvspace{-22pt}
\caption{\small{End-to-end per-step latency across six configurations (a)--(f). Each bar stacks the recompute stage over the policy update stage. Numbers above the bars denote the end-to-end speedup over veRL.}}
\label{fig:e2e}
\myvspace{-5pt}
\end{figure*}

\section{Evaluation}
\label{sec:eval}

Our evaluation is organized around four goals. We first quantify the end-to-end speedup that \system achieves over existing systems across different models, EP sizes, and datasets (\S\ref{subsec:eval_e2e}). We then decompose this speedup to isolate the contributions of individual planning stages and transfer-path choices (\S\ref{subsec:eval_ablation}). Next, we study how effectively \system reshapes rank load and inter-machine traffic, the two metrics targeted by our time model (\S\ref{subsec:eval_case}). Finally, we verify that both planning and expert transfer overheads can be overlapped with training, even when scaling up (\S\ref{subsec:eval_overhead}).

\subsection{Experimental Setup}
\label{subsec:eval_setup}

\mysubsubsection{Testbed and measurements.}
All experiments are conducted on a cluster of 8 machines, each equipped with 8 NVIDIA H20 GPUs connected via NVLink.
Following the predominant design in modern RL post-training systems~\cite{stream_rl, areal, llama_rl, roll_flash, asyncflow, psrl}, we adopt a disaggregated setup where rollout and training (recompute and policy update) are placed on separate GPUs.
In this setting, we observe that the rollout stage can actually be fully overlapped with training, and thus the end-to-end per-step latency reduces to the sum of recompute and policy update latencies.
We measure this combined training latency to quantify \system's speedup.\footnote{\system's design is orthogonal to the choice of RL architecture: it natively supports both disaggregating and colocating rollout and training. We use the disaggregated setup here as it is the more widely adopted configuration and provides a clean, training-focused comparison.}

\mysubsubsection{Models and datasets.} We evaluate two MoE models, Qwen3-30B-A3B and Qwen3.5-35B-A3B. For Qwen3-30B-A3B we test $\text{EP}=16$ and $\text{EP}=32$, and for Qwen3.5-35B-A3B we test $\text{EP}=32$, yielding three parallelism configurations in total. Table~\ref{tab:model_config} summarizes the settings of each. We run each configuration on two RL datasets: DAPO-Math-17k~\cite{dapo_dataset} and CodeForces~\cite{codeforces_dataset}. Every training step draws 512 samples with a prompt length of 2K and a response length of 8K. Each rank is provisioned with $N_r = 2$ redundant expert slots.

\mysubsubsection{Baselines.}
We compare \system with three baselines:
\begin{itemize}[leftmargin=*]
\item \textbf{veRL.} veRL~\cite{verl} is the state-of-the-art RL post-training system, with rollout--training disaggregation further enabled. It places experts sequentially across ranks in a fixed layout and performs no runtime load balancing.
\item \textbf{veRL+EPLB.} We augment veRL with EPLB~\cite{eplb}, a representative step-level load balancer designed for MoE pre-training by DeepSeek-V3~\cite{deepseek_v3}. EPLB relies on expert load statistics collected from the previous step and applies a greedy algorithm to determine expert relocation and replication for the current step.
\item \textbf{Oracle.} Oracle is a hypothetically balanced construct where every rank carries the same load, yielding a global lower bound on latency that is not physically realizable.
\end{itemize}


\begin{table}[t]
\centering
\caption{\small{\system speedup over baselines in different RL stages (``Rec.'' denotes recompute and ``Upd.'' denotes policy update).}}
\myvspace{-5pt}
\label{tab:foremoe}
\small
\begin{tabular}{lcccccc}
\toprule
\multirow{2.5}{*}{Config.} & \multicolumn{2}{c}{veRL} & \multicolumn{2}{c}{veRL+EPLB} & \multicolumn{2}{c}{Orcale} \\
\cmidrule(lr){2-3} \cmidrule(lr){4-5} \cmidrule(lr){6-7}
 & Rec. & Upd. & Rec. & Upd. & Rec. & Upd.  \\
\midrule
(a) & $1.23\times$ & $1.22\times$ & $1.20\times$ & $1.13\times$ & $0.95\times$ & $0.93\times$ \\
(b) & $1.55\times$ & $1.40\times$ & $1.31\times$ & $1.19\times$ & $0.96\times$ & $0.90\times$ \\
(c) & $1.35\times$ & $1.20\times$ & $1.28\times$ & $1.11\times$ & $0.96\times$ & $0.91\times$ \\
(d) & $1.24\times$ & $1.19\times$ & $1.15\times$ & $1.11\times$ & $0.96\times$ & $0.94\times$ \\
(e) & $1.47\times$ & $1.38\times$ & $1.26\times$ & $1.16\times$ & $0.96\times$ & $0.91\times$ \\
(f) & $1.35\times$ & $1.22\times$ & $1.25\times$ & $1.13\times$ & $0.97\times$ & $0.95\times$ \\
\bottomrule
\end{tabular}
\myvspace{-5pt}
\end{table}

\subsection{End-to-End Performance}
\label{subsec:eval_e2e}

Figure~\ref{fig:e2e} reports the per-step latency across six configurations, while Table~\ref{tab:foremoe} further breaks down the speedups of \system in each RL stage relative to the baselines.

Overall, \system accelerates per-step latency by $1.20$--$1.45\times$ over veRL and by $1.12$--$1.22\times$ over the step-level load balancing baseline veRL+EPLB. In the recompute stage, \system reduces latency by $1.23$--$1.55\times$ relative to veRL and by $1.15$--$1.31\times$ relative to veRL+EPLB. In the policy update stage, the corresponding reductions are $1.19$--$1.40\times$ and $1.11$--$1.19\times$, respectively. Furthermore, \system achieves $95$--$97\%$ of the Oracle speedup in the recompute stage and $90$--$95\%$ in the policy update stage. These results lead to three observations. First, while step-level load balancing can partially mitigate expert imbalance by addressing stable step-level skewness, a noticeable gap remains between veRL+EPLB and \system, highlighting the importance of micro-step-level load balancing in RL post-training. Second, while both stages benefit from micro-step-level load balancing, the gains are more pronounced during the recompute stage. As discussed in \S\ref{subsec:cpu_gpu_transfer}, this is because the CPU-assisted transfer path provides access to the entire CPU-resident expert pool. This expands the placement space significantly, enabling \system to generate more effective load-balancing plans.  
Third, the small gap between \system and Oracle indicates limited remaining optimization headroom. In particular, the recompute stage exhibits less than a $5\%$ gap across all configurations. This suggests that \system approaches the load-balancing upper bound in MoE RL post-training.  

\subsection{Ablation Study}
\label{subsec:eval_ablation}

We dissect \system along two axes, the planning algorithm of the \emph{Four-stage Planner} (\S\ref{sec:planner}) and the transfer path of the \emph{Expert Transfer Engine} (\S\ref{sec:system}), using Qwen3-30B-A3B with $\text{EP}=32$ on the DAPO-Math-17k dataset as a representative.

\begin{figure}[t]
\centering
\includegraphics[width=\columnwidth]{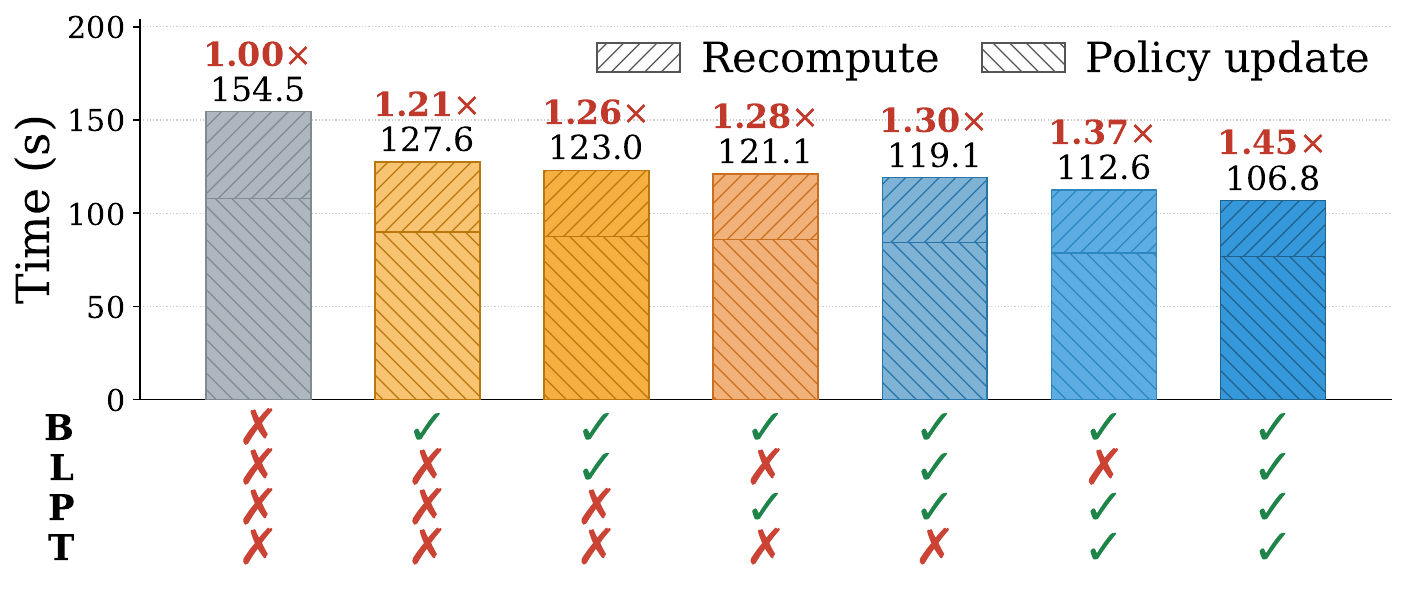}
\myvspace{-20pt}
\caption{\small{B, L, P, and T denote base expert placement, expert relocation, expert replication, and token assignment, respectively. The end-to-end speedup over veRL is shown on top.}}
\label{fig:ablation_algo}
\myvspace{-5pt}
\end{figure}

\mysubsubsection{Planning decomposition.} Figure~\ref{fig:ablation_algo} progressively enables the planner stages on top of veRL. The base mapping provides a large improvement ($1.21\times$) by removing the coarse step-level load skew. Relocation and replication then progressively refine per-rank balance, and the final token assignment among replicated experts further reduces the remaining imbalance. Combined, these stages deliver a $1.45\times$ end-to-end speedup. The results show that each stage contributes a non-overlapping latency reduction, validating the effectiveness of the four-stage decomposition.

\begin{table}[t]
\centering
\caption{\small{Transfer path comparison. We report the per-step latency of each RL stage under different expert transfer paths.}}
\myvspace{-5pt}
\label{tab:ablation_system}
\small
\begin{tabular}{lcc}
\toprule
Transfer path & Recompute & Policy update \\
\midrule
CPU-assisted       & \textbf{30.0s} & N/A \\
GPU-direct (intra-machine only)  & 34.9s & \textbf{76.8s} \\
GPU-direct  & 36.2s & 81.8s \\
\bottomrule
\end{tabular}
\myvspace{-10pt}
\end{table}

\mysubsubsection{Transfer path.} Table~\ref{tab:ablation_system} compares three transfer paths: CPU-assisted, intra-machine GPU-direct, and unrestricted GPU-direct. The first two correspond to the paths adopted by \system during recompute and policy update, respectively.

For the recompute stage, the CPU-assisted path achieves the best performance (30.0\,s). Keeping a complete expert pool in CPU exposes the full placement space to the planner, enabling more effective load balancing, while PCIe expert transfers are largely hidden via communication--computation overlap. In contrast, intra-machine GPU-direct reaches 34.9\,s, and unrestricted GPU-direct further degrades to 36.2\,s.

For the policy update stage, the CPU-assisted path is infeasible: maintaining gradient copies in host memory incurs prohibitive memory and synchronization overhead (see Appendix~\ref{app:cpu_pol}). Between the two GPU‑direct variants, the intra-machine design again proves superior, achieving 76.8\,s. 

Our evaluation of GPU-direct transfers reveals a clear trade-off. Although inter-machine transfers expand the placement space and yield a more balanced layout, realizing such layouts requires moving experts across the slower cross‑machine link. 
Unlike other paths, this communication cannot be effectively overlapped, so the transfer overhead outweighs the modest load‑balance gain. Thus, restricting GPU-direct transfers to intra‑machine communication yields better performance, validating the design rationale in \S\ref{subsec:cpu_gpu_transfer}.

\subsection{Case Study}
\label{subsec:eval_case}

\begin{figure}[t]
\centering
\includegraphics[width=\linewidth]{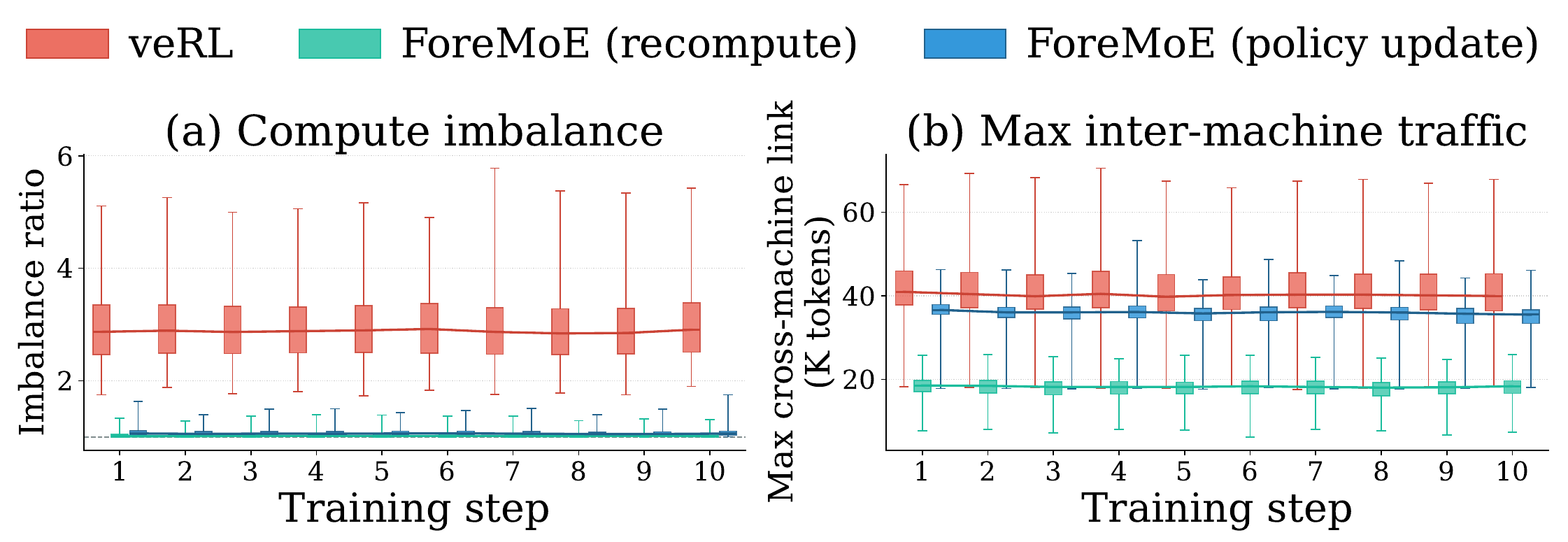}
\myvspace{-20pt}
\caption{\small{Per-step distribution of (a) the compute imbalance ratio and (b) the maximum inter-machine link traffic over 10 steps.}} 
\label{fig:casestudy}
\myvspace{-5pt}
\end{figure}

We next examine how \system reshapes the two key metrics in its time model (\S\ref{subsec:moe_time}, Eq.~(\ref{eq:layer_time_full})): the compute imbalance ratio $L_{\max}/\bar{L}$ and the maximum inter-machine link traffic $C_{\max}$. Figure~\ref{fig:casestudy} reports their distributions over 10 steps. For each step, every micro-step contributes one sample, and the resulting distribution is summarized as a box plot, with a line connecting the per-step medians.

For veRL, two observations emerge. First, both metrics remain highly stable across steps: neither the box distributions nor their medians shift noticeably over time. This validates the step-level stability assumption underlying \system's base placement (\S\ref{subsec:base_placement}). Second, substantial variation exists at the micro-step level within each step. Under natural placement, the imbalance ratio spans from below $2.5$ to nearly $5.8$, despite a median of only $2.9$. This pronounced volatility underscores the necessity of per-micro-step reconfiguration.

\system substantially reduces both metrics. During the recompute stage, where the planner can leverage the full expert pool, the median imbalance ratio drops to approximately $1.02$, while the median peak link traffic decreases from 40K to 18K tokens. 
During the policy update stage, despite being restricted to intra-machine transfers, the planner still reduces the median imbalance ratio to about $1.06$ and the median peak link traffic to approximately 36K tokens. 

\subsection{Overhead Analysis}
\label{subsec:eval_overhead}

In this section, we analyze the runtime overhead introduced by \system's planning (\S\ref{sec:planner}) and expert transfer (\S\ref{sec:system}). 

\begin{figure}[t]
\centering
\includegraphics[width=\columnwidth]{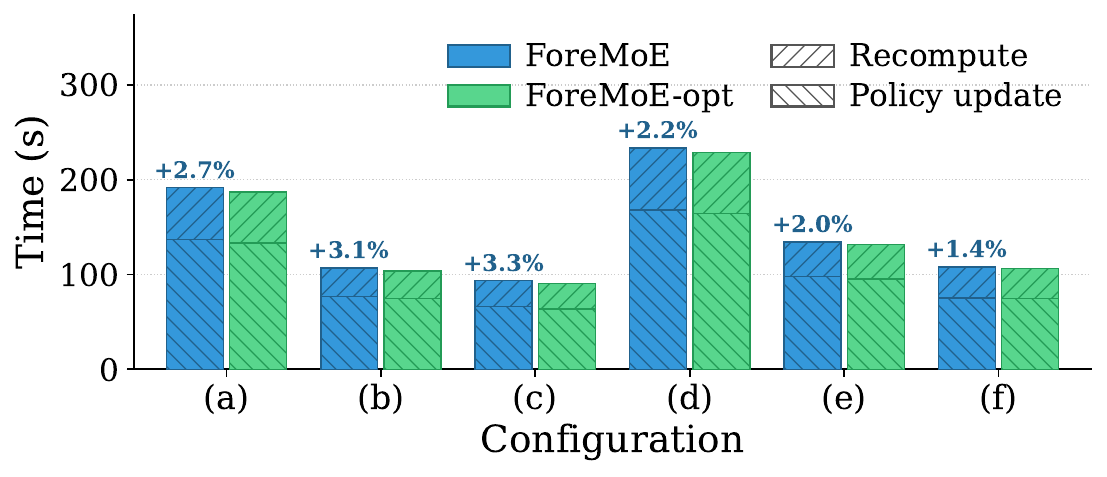}
\myvspace{-22pt}
\caption{\small{\system versus the idealized \system-opt.}}
\label{fig:opt_compare}
\myvspace{-5pt}
\end{figure}

\mysubsubsection{Comparison with \system-opt.}
We first compare \system with \system-opt, an idealized variant that executes planning and expert transfer offline, excluding all overheads to isolate the pure step latency achievable by our algorithm. This serves as an upper bound on \system's performance in the absence of planning and transfer costs. Figure~\ref{fig:opt_compare} reports the per-step latency under different configurations. The latency gap between \system and \system-opt remains within $1.4$--$3.3\%$, indicating that \system successfully overlaps nearly all planning and transfer overhead with training.

\mysubsubsection{Scalability.}
We next examine how the planning and expert-transfer overheads scale with cluster size. We fix $\text{EP}=16$ and increase the GPU count by scaling along the data parallelism. 

\begin{figure}[t]
\centering
\includegraphics[width=\linewidth]{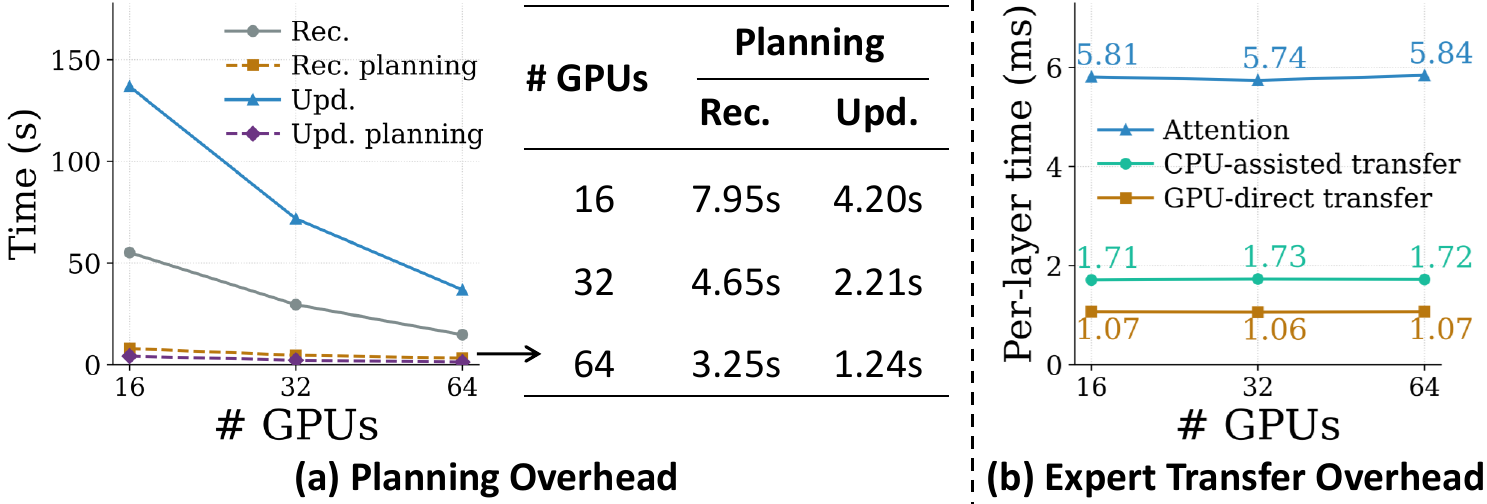}
\myvspace{-20pt}
\caption{\small{
(a) Per-step planning time versus stage time. ``Rec.'' denotes recompute and ``Upd.'' denotes policy update. (b) Per-layer expert transfer time versus attention time.}}
\label{fig:scalability}
\myvspace{-5pt}
\end{figure}

Figure~\ref{fig:scalability}(a) compares the planning time with the corresponding stage time, aggregated across all layers and micro-steps. As the cluster scales, micro-steps are distributed across a larger number of EP groups, reducing both the per-group workload and the corresponding planning time proportionally. Results show that planning remains a small fraction of training time at every scale, accounting for at most $22\%$ of the recompute stage and $3\%$ of the policy update stage. Such overhead can therefore be fully hidden through overlap. 

Figure~\ref{fig:scalability}(b) reports the per-layer expert transfer time, alongside the attention time it overlaps with. Because each EP group performs its relocation and replication independently, the transfer cost is invariant to the GPU count, holding at roughly 1.72\,ms for the CPU-assisted path and 1.07\,ms for the GPU-direct path (intra-machine) across all scales. Both remain well below the attention time of about 5.8\,ms. This confirms that the transfer overhead stays fully overlapped regardless of the cluster scale. 

\section{Related Work}
\label{sec:related}

\mysubsubsection{MoE load balancing.}
Mitigating expert load imbalance has been studied from both algorithmic and systems perspectives. Algorithmic methods include auxiliary balancing losses~\cite{loadloss, glam, switch_transformer}, expert capacity with token dropping~\cite{gshard, tutel}, and modified routing policies~\cite{base_layers, expert_choice, hash_layers}, which trade model quality for more uniform load. Systems approaches include runtime hot-expert replication~\cite{fastermoe}, periodic relocation based on popularity~\cite{smartmoe}, fine-grained replica adjustment~\cite{flexmoe}, decoupling expert parameters from optimizer states for per-iteration redistribution~\cite{symi}, fully-sharded expert parallelism with frequent re-layout hidden by prefetching~\cite{laer_moe}, and fast-paced token rescheduling~\cite{fine_moe}. 

These works all target pre-training, where load imbalance typically manifests at a coarse granularity. Hence, many methods reconfigure only at step boundaries~\cite{fastermoe, smartmoe, flexmoe, symi}, failing to capture the micro-step-level fluctuations that dominate RL post-training. Moreover, pre-training cannot obtain routing information ahead of time, so many methods rely on historical statistics~\cite{fastermoe, smartmoe, flexmoe, symi, popfetcher, laer_moe}. Acquiring exact loads online would otherwise compress the time budget for decision and reconfiguration on the critical path~\cite{fine_moe}. In RL post-training, however, routing is foreseeable. \system exploits this property to enable micro-step-level optimization.

Concurrent with our work, Relibra~\cite{relibra} also recognizes the importance of MoE load balancing in RL post-training. Similar to \system, it leverages rollout routing information and performs reconfiguration at micro-step granularity. However, Relibra focuses exclusively on the policy update stage and adopts only GPU-direct expert transfer. Moreover, while it supports expert replication, it does not support expert relocation as a micro-step-level reconfiguration mechanism. By contrast, \system supports both CPU-assisted and GPU-direct transfer paths, optimizes both the recompute and policy update stages, and jointly explores expert relocation and replication. Our ablation study (\S\ref{subsec:eval_ablation}) demonstrates that these additional capabilities are all important contributors to \system's performance gains, as exemplified by the gains from expert relocation in Figure~\ref{fig:ablation_algo} and the benefits of the CPU-assisted transfer path in Table~\ref{tab:ablation_system}.

\mysubsubsection{MoE efficiency optimization.}
Another line of work focuses on MoE efficiency optimization. Existing efforts either optimize communication, such as Tutel's topology-aware 2D Hierarchical All-to-All~\cite{tutel} and DeepEP's specialized All-to-All kernels~\cite{deepep}, or overlap communication with computation through expert-workload chunking~\cite{pipemoe, schemoe}. Comet~\cite{comet} further develops GPU kernel fusion via thread-block specialization. These techniques are orthogonal to \system, as they optimize MoE execution itself and can be directly combined with our load-balancing mechanisms.

\mysubsubsection{RL post-training systems.}
A growing body of work has explored efficient RL post-training for LLMs. Early systems such as OpenRLHF~\cite{openrlhf} adopt colocated architectures, while more recent designs embrace asynchronous and disaggregated execution~\cite{stream_rl, llama_rl, areal, roll_flash, rhy_rl, laminar, async_rlhf, asyncflow, psrl} to decouple rollout from training, improving resource utilization. 
Other works accelerate rollout through speculative decoding~\cite{seer, specrl, rhy_rl, respec, beat_long_tail} and low-precision quantization~\cite{qerl, roll_flash}. While these systems address utilization and efficiency, they do not explicitly tackle MoE load balancing. Thus, their optimization objectives are largely orthogonal to ours, allowing \system to be seamlessly integrated with these approaches.


\section{Conclusion}
\label{sec:conclusion}

This paper presents the first study of load imbalance characteristics in MoE RL post-training. We show that, unlike MoE pre-training where imbalance primarily occurs at step granularity, RL post-training exhibits substantial micro-step-level load fluctuations, making fine-grained balancing necessary. Motivated by this observation, we design \system, a load-balancing system that leverages foreseeable routing information from rollout as an accurate planning signal. \system enables micro-step-level load balancing through a four-stage planning process and  rapid reconfiguration via complementary expert transfer paths. Experiments on a 64-GPU cluster demonstrate that \system achieves up to $1.45\times$ end-to-end speedup over state-of-the-art MoE RL post-training systems.

\bibliographystyle{plain}
\bibliography{reference}

\clearpage

\appendix

\section{Detailed Overhead Analysis of the Expert Transfer Engine}
\label{app:overhead}

This appendix provides a formal overhead analysis of the two \emph{Expert Transfer Engine} designs described in \S\ref{sec:system}. We first establish a per-layer compute time model that expresses the GPU's available overlap budget in terms of model and workload variables (\S\ref{app:overhead_layer_model}). We then analyze the CPU-assisted design (\S\ref{app:overhead_cpu}) and the GPU-direct design (\S\ref{app:overhead_gpu}), each with explicit overlap conditions and scaling discussions. \S\ref{app:overhead_qwen3} concludes with a concrete instantiation for the Qwen3-30B-A3B model on our H20 platform as numerical confirmation.

For the analysis we consider one rank processing a single sequence of length $n$ in one micro-step. Throughout, let $E$ denote the number of experts per MoE layer, $K$ the number of experts activated per token (top-$K$), $h$ the hidden dimension, $h_{ff}$ the FFN intermediate dimension of one expert, and $L$ the number of MoE layers held by each training worker (after pipeline parallelism partitioning, if any). Let $S_e$ denote the size of one expert in the model's training precision and $S_g$ the size of one expert's gradient tensor (in the precision used for gradient accumulation). For a SwiGLU expert with three weight matrices (gate, up, down) of size $h\times h_{ff}$ each, $S_e = 3\,h\,h_{ff}\cdot p_w$ bytes where $p_w$ is the per-element parameter byte width, and $S_g$ has the same structure with the per-element gradient byte width $p_g$. Let $N_s$ denote the number of experts each rank holds in the placement (base plus redundant), $F$ the rank's effective compute throughput in FLOPS, $B_{\text{PCIe}}$ the achievable host-to-device PCIe bandwidth per GPU, and $B_{\text{NVLink}}$ the achievable per-GPU NVLink bandwidth.

\subsection{Per-Layer Compute Time Model}
\label{app:overhead_layer_model}

We model the per-layer compute time of one rank as the sum of the attention block and the MoE block. Following the standard FLOPs accounting for transformer layers~\cite{megatron_1, megatron_2}, the per-rank forward FLOPs of one layer are:
\begin{equation}
\small
\text{FLOPs}_{\text{attn}} \;=\; \underbrace{8\,n\,h^2}_{\text{QKV+output proj.}} \;+\; \underbrace{2\,n^2\,h}_{\text{causal attention}},
\label{eq:flops_attn}
\end{equation}
\begin{equation}
\small
\text{FLOPs}_{\text{moe}} \;=\; \underbrace{6\,K\,n\,h\,h_{ff}}_{\text{expert FFN (SwiGLU)}} \;+\; \underbrace{2\,n\,h\,E}_{\text{router}},
\label{eq:flops_moe}
\end{equation}
where the factor of $6$ in the FFN term reflects two FLOPs per multiply-add applied to the three weight matrices of a SwiGLU expert, and the FFN term assumes that the $K\,n$ token-expert pairs received by this rank under a balanced placement are processed locally. Dividing by the rank's effective throughput $F$ yields:
\begin{equation}
\small
T_{\text{attn}} \;=\; \frac{8\,n\,h^2 + 2\,n^2\,h}{F},\qquad
T_{\text{moe}} \;=\; \frac{6\,K\,n\,h\,h_{ff} + 2\,n\,h\,E}{F}.
\label{eq:t_attn_moe}
\end{equation}
The total compute time of one MoE layer on this rank is
\begin{equation}
\small
T_{\text{layer}} \;=\; T_{\text{attn}} + T_{\text{moe}} + T_{\text{a2a}},
\label{eq:t_layer}
\end{equation}
where $T_{\text{a2a}}$ accounts for the dispatch/combine All-to-All. For the overhead analyses below it suffices to lower-bound $T_{\text{layer}}\geq T_{\text{attn}}+T_{\text{moe}}$, and overlap budgets are taken against this conservative estimate.

The relative magnitudes of the FFN and attention terms are governed by
\begin{equation}
\small
\frac{T_{\text{moe}}}{T_{\text{attn}}} \;\approx\; \frac{6\,K\,h_{ff}}{8\,h + 2\,n},
\label{eq:moe_to_attn}
\end{equation}
which grows linearly in $K$ and $h_{ff}$ and shrinks as $n$ increases.

\subsection{CPU-assisted Expert Transfer}
\label{app:overhead_cpu}

We analyze the storage and transfer overheads of the CPU-assisted design used in the recompute stage.

\mysubsubsection{CPU memory.} Each training worker stores a master copy of every expert in every layer it owns, requiring $L \cdot E \cdot S_e$ bytes of pinned host memory. Because the recompute stage carries no gradient or optimizer state, this is the only expert-related memory placed on CPU, and modern host memory comfortably exceeds $L \cdot E \cdot S_e$ for the model sizes targeted by \system.

\mysubsubsection{GPU memory.} Each rank holds at most $N_s$ experts at a time. Incoming prefetches are written directly into the GPU memory regions that hold the rank's expert parameters, so no separate buffer is required on the GPU. The per-rank GPU footprint is $\Theta(N_s \cdot S_e)$, which matches the working set of a baseline EP layout with the same per-rank expert count. The CPU-assisted design therefore introduces no additional GPU memory overhead.

\mysubsubsection{Prefetch latency.} In the worst case, all $N_s$ experts on a rank are replaced between two consecutive micro-steps, generating $N_s \cdot S_e$ bytes of host-to-device traffic per rank per layer. The prefetch is fully hidden whenever
\begin{equation}
\small
T_{\text{prefetch}} \;=\; \frac{N_s \cdot S_e}{B_{\text{PCIe}}} \;\le\; T_{\text{layer}}.
\label{eq:cpu_overlap}
\end{equation}
Substituting Eqs.~(\ref{eq:t_attn_moe})--(\ref{eq:t_layer}), with $S_e = 3\,h\,h_{ff}\,p_w$, the inequality becomes a quadratic constraint on $n$:
\begin{equation}
\small
2\,n^2 + (8\,h + 6\,K\,h_{ff} + 2\,E)\,n \;\ge\; \frac{3\,N_s\,h_{ff}\,p_w \cdot F}{B_{\text{PCIe}}},
\label{eq:cpu_n_min}
\end{equation}
which is satisfied for all $n \ge n^{\text{cpu}}_{\min}$, where $n^{\text{cpu}}_{\min}$ is the positive root of the corresponding quadratic equation.

\mysubsubsection{Discussions.} Two observations follow from Eq.~(\ref{eq:cpu_n_min}). First, the LHS is strictly increasing in $n$ while the RHS is independent of $n$, so the inequality is always satisfied for sufficiently large sequences. The question is only whether $n^{\text{cpu}}_{\min}$ is small enough to be of practical concern. Second, $n^{\text{cpu}}_{\min}$ depends on the hardware through the ratio $F/B_{\text{PCIe}}$, the GPU's compute-density-per-byte. For modern accelerators this ratio is on the order of $10^3$ FLOPs/byte, while the LHS coefficient $8h + 6Kh_{ff} + 2E$ is on the order of $10^4$ to $10^5$ for production-scale MoE models, so $n^{\text{cpu}}_{\min}$ stays in the low thousands of tokens, which is small compared to a typical RL-post-training micro-step. Larger $h$, $h_{ff}$, $K$, or $E$ all increase the LHS coefficient and therefore shrink $n^{\text{cpu}}_{\min}$ further, so the bound holds increasingly comfortably for larger models.

\subsection{GPU-direct Expert Transfer (Intra-machine)}
\label{app:overhead_gpu}

We analyze the storage and transfer overheads of the GPU-direct design used in the policy update stage. Because each swapped expert must take its gradient with it (the gradient produced under the previous placement must continue to be accumulated against the same parameter on its new rank), every transfer in this stage carries both the expert's parameters and its gradients.

\mysubsubsection{GPU memory.} Each rank still holds at most $N_s$ experts. The transfer pipeline additionally allocates one send buffer and one receive buffer per rank, each sized to hold a rank's expert state (parameters and gradients of all $N_s$ experts on the rank), and shared across all MoE layers. The asymptotic GPU memory overhead is therefore $\Theta(N_s \cdot (S_e + S_g))$ on top of the placement working set.

\mysubsubsection{Swap latency.} Copy-out and Copy-in are intra-GPU memory operations whose cost is negligible compared to the MoE combine and dispatch they overlap with. The dominant stage is the NVLink swap, which carries up to $N_s \cdot (S_e + S_g)$ bytes per rank per layer and is overlapped with Attention. The swap is therefore fully hidden whenever
\begin{equation}
\small
T_{\text{swap}} \;=\; \frac{N_s\,(S_e + S_g)}{B_{\text{NVLink}}} \;\le\; T_{\text{attn}}.
\label{eq:gpu_overlap}
\end{equation}
Writing $S_e + S_g = 3\,h\,h_{ff}\,(p_w+p_g)$ and substituting Eq.~(\ref{eq:t_attn_moe}), the condition becomes
\begin{equation}
\small
2\,n^2 + 8\,h\,n \;\ge\; \frac{3\,N_s\,h_{ff}\,(p_w+p_g)\cdot F}{B_{\text{NVLink}}},
\label{eq:gpu_n_min}
\end{equation}
which is satisfied for all $n \ge n^{\text{nv}}_{\min}$, where $n^{\text{nv}}_{\min}$ is the positive root of the corresponding quadratic equation.

\mysubsubsection{GPU compute contention.} NVLink collectives are realized through GPU streaming multiprocessors that issue the underlying memory transactions, which introduces a small amount of contention with the compute stream during the swap. This contention is bounded by the small fraction of total layer time that the NVLink transfer occupies, and we measure it to be negligible in practice.

\mysubsubsection{Discussions.} Three observations follow from Eq.~(\ref{eq:gpu_n_min}). First, the relevant compute-density-per-byte is $F/B_{\text{NVLink}}$, which on modern intra-machine fabrics is roughly $5\text{--}10\times$ smaller than $F/B_{\text{PCIe}}$, so the high NVLink bandwidth dominates the gradient-inflated transfer volume. Second, the swap is hidden behind Attention alone (rather than the full $T_{\text{layer}}$), which is why Eq.~(\ref{eq:gpu_n_min}) lacks the FFN-related terms. In exchange, the LHS still grows quadratically in $n$ via the attention compute, so $n^{\text{nv}}_{\min}$ remains modest as long as the sequence length is not vanishingly small. Third, the linear factor $(p_w+p_g)$ exposes the cost of carrying gradients: training in pure BF16 ($p_g = p_w = 2$) doubles the swap volume relative to parameters alone, while FP32 gradient accumulation roughly triples it. Both regimes remain well within the $T_{\text{attn}}$ budget on modern NVLink fabrics, as the next subsection illustrates.

\subsection{Concrete Instantiation: Qwen3-30B-A3B on H20}
\label{app:overhead_qwen3}

We instantiate the bounds with the Qwen3-30B-A3B~\cite{qwen_3} model and our H20-based deployment. The model has $h=2048$, $h_{ff}=768$, $E=128$, $K=8$, and $48$ MoE layers. We assume BF16 weights ($p_w=2$) with FP32 gradient accumulation ($p_g=4$), so $S_e = 6\,h\,h_{ff} = 9.0$\,MiB and $S_g = 12\,h\,h_{ff} = 18.0$\,MiB per expert. Per H20 GPU, $F\approx 148\,\text{TFLOPS}$ in BF16, $B_{\text{PCIe}}\approx 64\,\text{GB/s}$ (PCIe Gen5 x16), and $B_{\text{NVLink}}\approx 450\,\text{GB/s}$.

We take $N_s = E/P + N_r = 128/16 + 2 = 10$ as a representative working set (EP=16 layout with two redundant slots per rank), and $n=8\,\text{K}$ as a representative sequence length for long-context RL post-training.

\mysubsubsection{CPU storage.} Per worker, $L\cdot E\cdot S_e = 48\cdot 128\cdot 9.0\,\text{MiB} = 54\,\text{GiB}$ of pinned host memory, well within the $\geq 1\,\text{TiB}$ host capacity of our machines.

\mysubsubsection{Prefetch budget (recompute).} Plugging the parameters above into Eq.~(\ref{eq:cpu_n_min}) gives $n^{\text{cpu}}_{\min} \approx 1862$ tokens. A single rank in our deployments routinely processes tens of thousands of tokens per micro-step, so this $\sim 1.9\,\text{K}$-token threshold is met by an order of magnitude or more. Equivalently, the worst-case prefetch volume of $N_s\cdot S_e = 90\,\text{MiB}$ takes $1.47\,\text{ms}$ over PCIe, while $T_{\text{layer}}$ for $n=8\,\text{K}$ tokens is $\approx 7.9\,\text{ms}$, leaving more than $6\,\text{ms}$ of headroom.

\mysubsubsection{Swap budget (policy update).} Plugging the parameters above into Eq.~(\ref{eq:gpu_n_min}) gives $n^{\text{nv}}_{\min} \approx 2189$ tokens, again well below the per-rank token count of a typical micro-step. Equivalently, the worst-case swap volume of $N_s\cdot(S_e+S_g) = 270\,\text{MiB}$ takes $0.63\,\text{ms}$ over NVLink, while $T_{\text{attn}}$ for $n=8\,\text{K}$ tokens is $\approx 3.7\,\text{ms}$, a $5.9\times$ margin.

These numerical margins confirm that the inequalities Eq.~(\ref{eq:cpu_overlap}) and Eq.~(\ref{eq:gpu_overlap}) hold by a comfortable factor in our deployment, and the scaling discussions of \S\ref{app:overhead_cpu} and \S\ref{app:overhead_gpu} indicate that the margins grow rather than shrink as models become deeper, wider, or longer-context.

\section{On a CPU-assisted Variant for the Policy Update Stage}
\label{app:cpu_pol}

The CPU-assisted design adopted for the recompute stage (\S\ref{app:overhead_cpu}) gains its expressiveness from holding the entire expert pool in pinned host memory and prefetching only what each micro-step requires. A natural question is whether the same approach extends to the policy update stage, where it would offer the same global placement flexibility. We have given this extension careful consideration. As we describe below, the bandwidth side admits a workable schedule (\S\ref{app:cpu_pol_bw}), but the memory side exposes a structural constraint specific to gradient transfer (\S\ref{app:cpu_pol_mem}) that ultimately drives our choice of the GPU-direct design as the default in \system.

\subsection{Bandwidth}
\label{app:cpu_pol_bw}

Extending the recompute-stage CPU-assisted design to policy update introduces a second entity beyond the parameters: the main gradients, which are roughly twice the size of the parameters under BF16 weights and FP32 gradient accumulation ($p_w=2$, $p_g=4$, hence $S_g=2\,S_e$), and which must additionally be written back to CPU after each backward layer's gradient accumulation completes.

We allocate the parameter and gradient traffic across the two passes as follows. The forward pass is kept identical to the recompute case: during the forward of layer $L$, prefetch the parameters of layer $L+1$ over PCIe on a separate CUDA stream, overlapping the on-going forward computation. The backward pass then carries all of the gradient traffic: during the backward of layer $L$, the transfer stream issues, in pipelined order, (1) a prefetch of the gradients of layer $L-1$ so that they are on GPU before $L-1$'s backward begins, and (2) an offload of the just-accumulated gradients of $L$ to CPU.

Per layer per rank, the resulting transfer volumes and overlap budgets are:
\begin{itemize}[leftmargin=*, noitemsep, topsep=0pt]
\item Forward: transfers $N_s\,S_e$ bytes, overlapped against $T_{\text{layer}}$ (identical to the recompute case).
\item Backward: transfers $2\,N_s\,S_g = 4\,N_s\,S_e$ bytes (one prefetch and one offload of the gradients), overlapped against backward compute of approximately $2\,T_{\text{layer}}$.
\end{itemize}
The transfer-to-compute ratio is $N_s S_e/(B_{\text{PCIe}}\,T_{\text{layer}})$ on the forward side and $2\,N_s S_e/(B_{\text{PCIe}}\,T_{\text{layer}})$ on the backward side. Therefore backward is the binding constraint, with overlap budget approximately twice as tight as in the recompute case (e.g., on Qwen3-30B-A3B + H20 the binding margin shrinks from $\sim\!5.4\times$ to $\sim\!2.7\times$).

\subsection{Memory: A Structural Constraint on Gradient Sharing}
\label{app:cpu_pol_mem}

Parameters and gradients are not symmetric in how they consume memory resource. The parameters are read-only and identical across workers, so a single shared memory-mapped region per machine suffices, leaving the pinned-memory footprint unchanged from the recompute case ($L\cdot E\cdot S_e$ per machine). The gradients, by contrast, are write-shared via expert replication: when an expert $e$ is replicated on multiple ranks of the same machine, each rank's backward pass independently accumulates into "the gradient of $e$", so a shared backing buffer would invite concurrent writes that demand atomic accumulation or explicit cross-process locking, neither of which is cheap to implement in pinned host memory accessed over PCIe.

The conservative alternative is to give every worker its own private gradient master copy, at $L\cdot E\cdot S_g$ pinned bytes per worker. For Qwen3-30B-A3B this is $48\cdot 128\cdot 18\,\text{MiB}=108\,\text{GiB}$ per worker, which on an 8-GPU machine sums to $\approx 0.85\,\text{TiB}$. For Qwen3-235B-A22B without pipeline parallelism it climbs into the multi-TiB range per machine. Moreover, pinned memory differs from ordinary host memory in that it is non-pageable kernel-managed memory and competes directly with the OS file cache and other process allocations. Committing it at this scale degrades machine-wide throughput well before nominal RAM is exhausted, and risks OOM under memory-pressure spikes.

\subsection{Outlook}
\label{app:cpu_pol_outlook}

The schedule of \S\ref{app:cpu_pol_bw} tightens the overlap requirement above that of the recompute case, and the memory analysis of \S\ref{app:cpu_pol_mem} shows that a per-worker gradient master copy at production model scale exceeds the practical pinned-memory budget of present-day machines. \system\ therefore adopts the GPU-direct design of \S\ref{app:overhead_gpu} as its policy-update transfer engine. The CPU-assisted variant nonetheless remains an attractive direction in hardware regimes with abundant pinned host memory and lightweight cross-worker gradient-synchronization mechanisms, and we leave a careful exploration to future work.

\section{MILP Constraint Details}
\label{app:milp_constraints}

This appendix expands on the four constraints used in the joint MILP formulation of \S\ref{subsec:milp}, explaining the structural property each one enforces.

\mysubsubsection{Expert slot capacity ($\sum_{e} x_{e,j} = 1$).} A slot is a single physical position on a rank that can hold one expert tensor, so two experts cannot occupy the same slot.

\mysubsubsection{Expert coverage ($\sum_{j} x_{e,j} \geq 1$).} Every expert in the model must be placed somewhere; otherwise its tokens have no destination. The inequality (rather than equality) allows replicas: an expert may be placed in more than one slot whenever the load profile justifies replication.

\mysubsubsection{Token conservation ($\sum_{j} r_{s,e,j} = 1$).} For every (source rank, expert) pair, the entire token volume must be dispatched to some destination slot. This forbids dropping tokens. When expert $e$ has multiple replicas, this constraint allows source $s$'s tokens to be assigned among them in any proportions that sum to one.

\mysubsubsection{Token assignment feasibility ($r_{s,e,j} \leq x_{e,j}$).} Tokens for expert $e$ may travel only to slots that actually hold $e$. If expert $e$ is not in slot $j$ (so $x_{e,j} = 0$), this constraint forces $r_{s,e,j} = 0$, preventing any tokens from being routed to a slot that does not host the corresponding expert.

\section{Policy Update Stage Algorithm}
\label{app:update_alg}

The policy update stage uses the GPU-direct exchange within a machine of \S\ref{subsec:cpu_gpu_transfer}, which carries both parameters and gradients per expert and confines relocation and replication to intra-machine NVLink transfers. In principle, Stages 2-4 of Algorithm~\ref{alg:per_micro_step} can be reused here almost unchanged, the only modification being an intra-machine restriction that rejects any swap or replica placement requiring a cross-machine transfer. However, we observe that once the search space is restricted to a single machine, the problem becomes considerably simpler, and a lighter-weight procedure with lower complexity attains the same balancing quality as the restricted form of Algorithm~\ref{alg:per_micro_step}. We therefore adopt this simpler procedure, given in Algorithm~\ref{alg:update}.

We still reuse the base placement $A_{\text{base}}$ from Stage 1 as the starting point of every micro-step. We then apply the simpler Stages 2-4 to refine the per-micro-step placement within each machine. Each of Stages 2-4 decomposes into $M$ independent per-machine subproblems, which are solved in parallel. We now describe these Stages 2-4 in detail.

\begin{algorithm}[t]
\caption{\small{Policy Update Stage Planner (Stages 2-4)}}
\label{alg:update}
\small
\begin{algorithmic}[1]
\Require Base placement $A_{\text{base}}$, per-micro-step loads $\{W^{(i)}\}_{i=1..N}$, machines $M$, ranks per machine $R$, redundant slots per rank $N_r$
\Ensure Per-micro-step placements $\{A^{(i)}\}$, assignments $\{r^{(i)}\}$
\For{each micro-step $i = 1..N$ \textbf{in parallel}}
    \State $A^{(i)} \gets A_{\text{base}}$
    \For{each machine $m = 1..M$ \textbf{in parallel}}
        \State \textbf{// Stage 2: intra-machine relocation via LPT}
        \State $\text{local\_experts} \gets \{e : \text{machine}(A^{(i)}[e]) = m\}$
        \State Sort $\text{local\_experts}$ by descending $\sum_s w^{(i)}_{s,e}$
        \State $\text{RL}[r] \gets 0$ for $r \in \text{ranks}(m)$
        \For{each $e$ in sorted $\text{local\_experts}$}
            \State $r^* \gets \arg\min_{r \in \text{ranks}(m)} \text{RL}[r]$
            \State $A^{(i)}[e] \gets r^*$; $\text{RL}[r^*] \mathrel{+}= \sum_s w^{(i)}_{s,e}$
        \EndFor
        \State \textbf{// Stage 3: intra-machine replication ($R \cdot N_r$ slots)}
        \For{$k = 1$ to $R \cdot N_r$}
            \State $e^* \gets$ locally heaviest expert
            \State $r^* \gets \arg\min_{r \in \text{ranks}(m)} \text{RL}[r]$
            \State Place a replica of $e^*$ on $r^*$ and update $\text{RL}$
        \EndFor
        \State \textbf{// Stage 4: water-filling token assignment}
        \For{each replicated expert $e$ on $m$}
            \State Assign each token unit to the replica whose rank has the smallest accumulated load
        \EndFor
    \EndFor
\EndFor
\State \Return $\{A^{(i)}\}, \{r^{(i)}\}$
\end{algorithmic}
\end{algorithm}

\mysubsubsection{Stage 2: intra-machine relocation.} For each machine, sort its hosted experts in descending order of micro-step-level load $\sum_s w^{(i)}_{s,e}$ and redistribute them across the machine's local ranks via LPT. This rebalances per-rank computation load within the machine.

\mysubsubsection{Stage 3: intra-machine replication.} For each machine, fill its local redundant slots one at a time. At each step, choose the locally heaviest expert and place its replica on the local rank with the minimum current load.

\mysubsubsection{Stage 4: water-filling token assignment.} Within each machine, distribute tokens for each replicated expert to its replicas by water-filling: iterate over source-rank tokens and assign each unit to the replica whose hosting rank currently has the minimum accumulated load, until the replicas' ranks converge toward equal utilization. We use water-filling rather than the LP of Eq.~(\ref{eq:lp_token}) because intra-machine replicas do not affect cross-machine communication, leaving the LP's joint computation-communication objective with little additional optimization space over the simpler heuristic.


\end{document}